\documentclass[11pt,aps,showpacs,nofootinbib,prd,aps,epsf,floats,
               amsmath,amssymb,amsfonts]{revtex4}
\usepackage{amsmath, amssymb}
\newcommand{\mathsym}[1]{{}}

\usepackage{graphicx}
\usepackage{amsmath}
\usepackage{amssymb}
\usepackage{bm}
\newcommand{\be}{\begin{equation}}
\newcommand{\ee}{\end{equation}}
\newcommand{\bea}{\begin{eqnarray}}
\newcommand{\eea}{\end{eqnarray}}

\newcommand{\rem}[1]{}
\newsavebox{\PSLASH}
 \sbox{\PSLASH}{$p$\hspace{-1.8mm}/}
 
\renewcommand{\theequation}{\thesection.\arabic{equation}}
\newcounter{saveeqn}
\newcommand{\add}{\addtocounter{equation}{1}}
\newcommand{\alpheqn}{\setcounter{saveeqn}{\value{equation}}%
\setcounter{equation}{0}%
\renewcommand{\theequation}{\mbox{\thesection.\arabic{saveeqn}{\alph{equation}}}}}
\newcommand{\reseteqn}{\setcounter{equation}{\value{saveeqn}}%
\renewcommand{\theequation}{\thesection.\arabic{equation}}}

 \newsavebox{\notrightarrow}
 \sbox{\notrightarrow}{$\to$\hspace{-4mm}/}
 
 \newsavebox{\PARTIALSLASH}
 \sbox{\PARTIALSLASH}{$\partial$\hspace{-1.6mm}/}
 
 \newsavebox{\ASLASH}
 \sbox{\ASLASH}{$A$\hspace{-2.1mm}/}
 
 \newsavebox{\KSLASH}
 \sbox{\KSLASH}{$k$\hspace{-1.8mm}/}
 
 \newsavebox{\LSLASH}
 \sbox{\LSLASH}{$\ell$\hspace{-1.8mm}/}
 
 \newsavebox{\QSLASH}
 \sbox{\QSLASH}{$q$\hspace{-1.8mm}/}
 
 \newsavebox{\DSLASH}
 \sbox{\DSLASH}{$D$\hspace{-2.2mm}/}
 
 \newsavebox{\DbfSLASH}
 \sbox{\DbfSLASH}{${\mathbf D}$\hspace{-2.8mm}/}
 
 \newsavebox{\DELVECRIGHT}
 \sbox{\DELVECRIGHT}{$\stackrel{\rightarrow}{\partial}$}
 
 \newcommand{\blue}{\IfColor{\textCadetBlue}{}}
\newcommand{\black}{\IfColor{\textBlack}{}}
\newcommand{\red}{\IfColor{\textRed}{}}
\newcommand{\green}{\IfColor{\textOliveGreen}{}}
\newcommand{\lila}{\IfColor{\textRedViolet}{}}

\begin{document}
\begin{flushright}
 0911.3461 [hep-th]
\end{flushright}
\title{Perturbative quantization of two-dimensional space-time noncommutative QED}
\author{M. Ghasemkhani}\email{ghasemkhani@physics.sharif.ir}
\author{N. Sadooghi}\email{sadooghi@physics.sharif.ir}
\affiliation{Department of Physics, Sharif University of Technology,
P.O. Box 11155-9161, Tehran-Iran}
\begin{abstract}
Using the method of perturbative quantization in the first order
approximation, we quantize a non-local QED-like theory including
fermions and bosons whose interactions are described by terms
containing higher order space-time derivatives. As an example, the
two-dimensional space-time noncommutative QED (NC-QED) is quantized
perturbatively up to ${\cal{O}}(e^{2},\theta^{3})$, where $e$ is the
NC-QED coupling constant and $\theta$ is the noncommutativity
parameter. The resulting modified Lagrangian density is shown to
include terms consisting of first order time-derivative and higher
order space-derivatives of the modified field variables that satisfy
the ordinary equal-time commutation relations up to
${\cal{O}}(e^{2},\theta^{3})$. Using these commutation relations,
the canonical current algebra of the modified theory is also
derived.
\end{abstract}
 \maketitle
\section{Introduction}\label{introduction}
\par\noindent
Theories including higher order time derivatives appears in
different areas of physics. In general relativity, for instance, the
quantum corrections added to the original lower derivative theory
contains higher derivatives of the metric \cite{birrell}. This
occurs also in the case of cosmic strings \cite{cosmic}, and in
Dirac's relativistic model of classical radiating electron
\cite{dirac, simon}. In contrast to the naive expectation, the
presence of an unconstrained higher-derivative term, no matter how
small it may appear, makes the new theory dramatically different
from its original lower-derivative counterpart. As it is shown in
\cite{simon}, unconstrained higher-derivative theories have more
degrees of freedom than lower derivative theories, and they lack of
a lower-energy bound. Classically, there is a no-go theorem by
Ostrogradski \cite{ostro}, who essentially showed why no more than
two time derivatives of the fundamental dynamical variables appear
in the laws of physics. Ostrogradski's result is that there is a
linear instability in the Hamiltonian associated with Lagrangians
which depend upon more than one time derivative in such a way that
the dependence cannot be eliminated by partial integration
\cite{ostro, woodard} (for a modern review of higher derivative
theories see \cite{woodard}).
\par
There is also a large class of theories containing higher
derivatives that do not suffer the above problems. Nonlocal
theories, where the nonlocality is regulated by a natural small
parameter contain in general implicit constraints which keep the
number of degrees of freedom constant and maintain a lower-energy
bound \cite{simon}. Nonlocality naturally appears in effective
theories in a low-energy limit that are derived from a larger theory
with some degrees of freedom frozen out \cite{simon}. A good example
is Wheeler-Feynman electrodynamics \cite{wheeler}, in which the
degrees of freedom of the electromagnetic field are frozen out
\cite{simon}. Space-time noncommutative field theories that arise
from open string in a background \textit{electric}
field,\footnote{Note that space-space noncommutative field theory
describes low energy limit of string theory in a background
\textit{magnetic} field.} is another example of nonlocal low energy
effective field theories consisting of an infinite number of
temporal and spatial derivatives in the interaction part of the
corresponding Lagrangian densities. The embedding of noncommutative
field theories into string theory is maybe relevant to understanding
the inevitable breakdown of our familiar notions of space and time
at short distances in quantum gravity \cite{noncommutative,
gopakumar}. Whereas space-space noncommutative theories suffer from
a mixing of ultraviolet and infrared singularities in their
perturbative dynamics \cite{minwalla}, the space-time noncommutative
theories seem to be seriously acausal and inconsistent with
conventional Hamiltonian evolution \cite{susskind}. Besides they do
not have a unitary S-matrix \cite{mehen}. Indeed the breakdown of
unitarity in a theory consisting of higher order time-derivative and
the above mentioned Ostrogradskian instability are closely related
\cite{woodard}. However, as it is shown in \cite{fredenhagen}, the
unitarity of the space-time noncommutative theories can be restored
and the path integral quantization can be performed. This progress
suggests that space-time noncommutative theories can be incorporated
in the framework of canonical quantization \cite{vassili}. Different
canonical approaches are suggested in \cite{gomis, vassili}. In
\cite{gomis}, first a general Hamiltonian formalism is developed for
nonlocal field theories in $d$ space-time dimensions by considering
auxiliary $d+1$ dimensional field theories which are local with
respect to the evolution in time. The case of noncommutative
$\varphi^{3}$ theory is then considered as an example. In
\cite{vassili}, a modification of the Poisson bracket suitable for a
canonical analysis of space-time noncommutative field theories is
constructed.
\par
Another possibility to quantize the space-time noncommutative gauge
theories is to use the perturbative quantization introduced in
\cite{ho-1,ho-2, eliezer}. In \cite{ho-1, ho-2}, the method of
perturbative quantization is used to define the Poisson structure
and Hamiltonian of generic higher derivative classical and quantum
field theories. This method is independently developed in
\cite{eliezer}. In \cite{ho-1}, the perturbative quantization of
noncommutative gauge theories is discussed qualitatively, as an
example, but no explicit calculation is performed. In \cite{ho-2},
the same method is used to quantize the Lagrange function involving
higher order time derivatives for both bosons and fermions in $0+1$
dimensions. As an example the supersymmetric noncommutative
Wess-Zumino model is considered. In all these examples higher order
time derivatives appear in the interaction part of the Lagrangian.
Recently, this method is also used in a series of paper by Reyes et
al. \cite{reyes-1, reyes-2} to quantize specific models, where the
nonlocal higher time derivative terms does not appear in the
interaction part of the Lagrangian density.
\par
The aim of the present paper is to quantize a two-dimensional
noncommutative QED perturbatively in the first order approximation
using the method described in \cite{ho-1,ho-2, eliezer}. The
resulting effective Lagrangian density of the theory will be also
presented in terms of modified field variables that satisfy the
ordinary equal-time commutation relations order by order in
perturbation theory. The organization of the paper is as follows. In
Sect. II, we will develop the general framework of perturbative
quantization for a $D+1$ dimensional QED-like theory including
bosons and fermions whose interactions are described by terms
containing higher order space-time derivatives. In Section III,
after introducing the noncommutative Moyal product, involving an
infinite number of space-time derivatives, we will quantize $1+1$
dimensional space-time noncommutative theory perturbatively up to
${\cal{O}}(e^{2},\theta^{3})$, where $e$ is the NC-QED coupling
constant and $\theta$ is the noncommutativity parameter. The
effective Lagrangian density of the theory including first order
time derivative and higher order space-derivatives is presented in
Sect. III. It includes a bosonic and a fermionic part. Whereas the
fermionic part is modified in this order of expansion, the bosonic
part remains unchanged. In Sect. IV, using the Dirac brackets of the
modified fields, we will determine, as a by product, the canonical
algebra of the global NC-$U_{V}(1)$ vector currents of the original
noncommutative theory up to ${\cal{O}}(e^{2},\theta^{3})$. Section
IV is devoted to discussions.
\section{Canonical Quantization of modified QED including higher order time derivatives}
\setcounter{equation}{0} \par\noindent Let us consider the
Lagrangian density of modified QED including higher order space-time
derivatives of fermionic and bosonic degree of freedom
$\psi,\bar{\psi}$ and $A_{\mu}$
\begin{eqnarray}\label{U1}
{\cal{L}}={\cal{L}}_{\mbox{\tiny{kin}}}
+{\cal{L}}_{\mbox{\tiny{int}}}(\psi,\partial_{\mu}\psi,\partial_{\mu}\partial_{\nu}\psi,\cdots;\bar\psi,\partial_{\mu}\bar\psi,\partial_{\mu}\partial_{\nu}\bar\psi,\cdots;A_{\rho},\partial_{\mu}A_{\rho},
\partial_{\mu}\partial_{\nu}A_{\rho},\cdots).
\end{eqnarray}
Here, we have assumed that higher order space-time derivatives
appear only in the interaction part,
${\cal{L}}_{\mbox{\tiny{int}}}$. The kinetic term,
${\cal{L}}_{\mbox{\tiny{kin}}}$, is therefore the ordinary kinetic
Lagrangian density of free QED in $D+1$ dimensions
\begin{eqnarray}\label{U2}
{\cal{L}}_{\mbox{\tiny{kin}}}=\bar{\psi}i\gamma^{\mu}\partial_{\mu}\psi-\frac{1}{4}
{\cal{F}}_{\mu\nu}{\cal{F}}^{\mu\nu}-\frac{1}{2\xi}(\partial_{\mu}
A^{\mu})^{2},
\end{eqnarray}
where ${\cal{F}}_{\mu\nu}\equiv
\partial_{\mu}A_{\nu}-\partial_{\nu}A_{\mu}$ is the field
strength tensor of ordinary QED and $\xi$ is the gauge fixing
parameter. In this section, we will perturbatively quantize the
theory described by (\ref{U1}) up to order ${\cal{O}}(e^{2})$, where
$e$ is the coupling constant of bosons and fermions. To this
purpose, we will first introduce the corresponding fermionic and
bosonic symplectic two-forms, from which the non-trivial Poisson
algebra of these fields can be derived. After replacing higher order
time derivatives with the corresponding space-derivatives using the
Euler-Lagrange equation of motion arising from (\ref{U1}) up to
${\cal{O}}(e)$, the field variables $\bar{\psi},\psi$, and $A_{\mu}$
will be appropriately redefined so that the modified field variables
satisfy the ordinary fundamental Poisson brackets. The resulting
Poisson algebra will be then quantized using the well-known Dirac
quantization prescription.
\subsection{General Structure of Symplectic Two-Forms and Poisson Brackets}
\par\noindent
Let us start by varying the action $S=\int d^{D}x\ dt\
{\cal{L}}$, with ${\cal{L}}$ from (\ref{U1}),
\begin{eqnarray}\label{U3}
\delta S&=&\int
d^{D}x~dt\partial_{\nu_{1}}\left(\sum_{m,k=0}^{\infty}(-1)^{k}\partial_{\mu_{1}}\cdots\partial_{\mu_{k}}\frac{\partial{\cal{L}}}
{\partial\left(\partial_{\mu_{1}}\cdots\partial_{\mu_{k}}\partial_{\nu_{1}}\partial_{\nu_{2}}\cdots\partial_{\nu_{m+1}}\psi\right)}\right)\delta
\left(\partial_{\nu_{2}}\cdots\partial_{\nu_{m+1}}\psi\right)\nonumber\\
&&+\int
d^{D}x~dt\delta\left(\partial_{\nu_{2}}\cdots\partial_{\nu_{m+1}}\bar{\psi}\right)\partial_{\nu_{1}}\left(\sum_{m,k=0}^{\infty}(-1)^{k}\partial_{\mu_{1}}\cdots\partial_{\mu_{k}}\frac{\partial{\cal{L}}}
{\partial\left(\partial_{\mu_{1}}\cdots\partial_{\mu_{k}}\partial_{\nu_{1}}\partial_{\nu_{2}}\cdots\partial_{\nu_{m+1}}
\bar{\psi}\right)}\right)
\nonumber\\
&&+\int
d^{D}x~dt\partial_{\nu_{1}}\left(\sum_{m,k=0}^{\infty}(-1)^{k}\partial_{\mu_{1}}\cdots\partial_{\mu_{k}}\frac{\partial{\cal{L}}}
{\partial\left(\partial_{\mu_{1}}\cdots\partial_{\mu_{k}}\partial_{\nu_{1}}\partial_{\nu_{2}}\cdots\partial_{\nu_{m+1}}A_{\sigma}\right)}\right)\delta
\left(\partial_{\nu_{2}}\cdots\partial_{\nu_{m+1}}A_{\sigma}\right)
\nonumber\\
&&+\mbox{EoM},
\end{eqnarray}
where EoM is the space-time integral of the Euler-Lagrange equation
of motions
\begin{eqnarray}\label{U4}
\sum_{k=0}^{\infty}(-1)^{k}\partial_{\mu_{1}}\cdots\partial_{\mu_{k}}\frac{\partial{\cal{L}}}
{\partial\left(\partial_{\mu_{1}}\cdots\partial_{\mu_{k}}\psi\right)}&=&0,\nonumber\\
\sum_{k=0}^{\infty}(-1)^{k}\partial_{\mu_{1}}\cdots\partial_{\mu_{k}}\frac{\partial{\cal{L}}}
{\partial\left(\partial_{\mu_{1}}\cdots\partial_{\mu_{k}}\bar{\psi}\right)}&=&0,\nonumber\\
\sum_{k=0}^{\infty}(-1)^{k}\partial_{\mu_{1}}\cdots\partial_{\mu_{k}}\frac{\partial{\cal{L}}}
{\partial\left(\partial_{\mu_{1}}\cdots\partial_{\mu_{k}}A_{\sigma}\right)}&=&0,
\end{eqnarray}
multiplied by $\delta\psi$ from right, $\delta\bar{\psi}$ from left,
and $\delta A_{\mu}$, respectively. Neglecting the surface terms
with respect to spatial coordinates, the remaining terms in
(\ref{U3}) can be written as
\begin{eqnarray}\label{U5}
\delta S&=&\int d^{D}x~ dt~\sum_{m=0}^{\infty}
\partial_{0}\left(\Pi_{\psi^{(m)}}\delta
\psi^{(m)}-\delta\bar{\psi}^{(m)}\Pi_{\bar{\psi}^{(m)}}+\Pi_{A^{(m)}}^{\sigma}\delta
A^{(m)}_{\sigma}\right).
\end{eqnarray}
Here, the superscripts $(m)$ denote the $m$-th order time derivative
of the corresponding fields, and the canonical momenta corresponding
to fermions $\Pi_{\psi^{(m)}}, \Pi_{\bar{\psi}^{(m)}}$, and to
bosons $\Pi_{A^{(m)}}^{\sigma}$ are given by
\begin{eqnarray}\label{U6}
\Pi_{\psi^{(m)}}&=&
\sum_{k=0}^{\infty}(-1)^{k}\partial_{\mu_{1}}\cdots\partial_{\mu_{k}}\frac{\partial^{R}{\cal{L}}}{\partial\left(\partial_{\mu_{1}}\cdots\partial_{\mu_k}\psi^{(m+1)}\right)},\nonumber\\
\Pi_{\bar{\psi}^{(m)}}&=&
\sum_{k=0}^{\infty}(-1)^{k}\partial_{\mu_{1}}\cdots\partial_{\mu_{k}}\frac{\partial^{L}{\cal{L}}}{\partial\left(\partial_{\mu_{1}}\cdots\partial_{\mu_k}\bar{\psi}^{(m+1)}\right)},\nonumber\\
\Pi_{A^{(m)}}^{\sigma}&=&
\sum_{k=0}^{\infty}(-1)^{k}\partial_{\mu_{1}}\cdots\partial_{\mu_{k}}
\frac{\partial{\cal{L}}}{\partial\left(\partial_{\mu_{1}}\cdots
\partial_{\mu_k}A^{(m+1)}_{\sigma}\right)}.
\end{eqnarray}
Here, $\partial^{R}$ and $\partial^{L}$ are the right and left
derivatives, respectively. Using (\ref{U5}) the symplectic two-form
$\Omega(t)$ is defined as\footnote{The relative minus sign between
$X^{(m)}$ and $Y^{(m)}$ appears also in \cite{villi-02}.}
\begin{eqnarray}\label{U7}
\Omega(t)=\sum_{m=0}^{\infty}\int d ^ {D}x
\left(X^{(m)}(t;{\mathbf{x}})-Y^{(m)}(t;{\mathbf{x}})+Z^{(m)}(t;{\mathbf{x}})\right),
\end{eqnarray}
where
\begin{eqnarray}\label{U8}
X^{(m)}(t;{\mathbf{x}})&=&d\Pi_{\psi^{(m)}}^{\beta}(t;{\mathbf{x}})\wedge
d\psi_{\beta}^{(m)}(t;{\mathbf{x}}),\nonumber\\
Y^{(m)}(t;{\mathbf{x}})&=&d\bar\psi_{\alpha}^{(m)}(t;{\mathbf{x}})\wedge
d\Pi_{\bar{\psi}^{(m)}}^{\alpha}(t;{\mathbf{x}}),\nonumber\\
Z^{(m)}(t;{\mathbf{x}})&=&
d\Pi^{\mu}_{A^{(m)}}(t;{\mathbf{x}})\wedge
dA_{\mu}^{(m)}(t;{\mathbf{x}}).
\end{eqnarray}
To determine the Poisson brackets of fermionic and bosonic degrees
of freedom, we will introduce the following equivalent
representation of the symplectic two-form $\Omega(t)$ from
(\ref{U7}),
\begin{eqnarray}\label{U9}
\Omega(t)=\Omega_{g}(t)+\Omega_{f}(t),
\end{eqnarray}
where $\Omega_{g}(t), \Omega_{f}(t)$ are the gauge and the fermionic
parts of $\Omega(t)$, respectively. The gauge part is given by
\begin{eqnarray}\label{U10}
\Omega_{g}(t)\equiv\frac{1}{2}\int d^{D}x\ d^{D}x'~
W_{ab}(t;{\mathbf{x}},{\mathbf{x}}')~dz^{a}(t;{\mathbf{x}})\wedge
dz^{b}(t;{\mathbf{x}}'),
\end{eqnarray}
where the phase space variables are $(D+1)(m+1)$-dimensional vector
$$z^{a}\equiv (A^{0},\cdots
A^{D},\dot{A}^{0},\cdots,\dot{A}^{D},\cdots A^{0(m)}, \cdots,
A^{D(m)},\cdots),$$ with $m\in \{0,\cdots,\infty\}$ determining the
order of time derivatives acting on $A_{\mu}$, with $\mu=0,1,\cdots
D$. In (\ref{U10}), the operator $W_{ab}$ is determined using
$Z^{(m)}(\textbf{x})$ from (\ref{U8}). Using $\Omega_{g}(t)$ from
(\ref{U10}), the Poisson bracket between $z^{a}$ is defined as
\begin{eqnarray}\label{U11}
\{z^{a}(t;{\mathbf{x}}),z^{b}(t;{\mathbf{x}}')\}_{\mbox{\tiny{PB}}}=W^{ab}(t;{\mathbf{x}},{\mathbf{x}}'),
\end{eqnarray}
where $W^{ab}(t;x,x')$ is the inverse of the operator
$W_{ab}(t;x,x')$ appearing in (\ref{U10}) and satisfies
\begin{eqnarray}\label{U11-b}
W^{ab}(t;{\mathbf{x}},{\mathbf{x}}')=-W^{ba}(t;{\mathbf{x}}',{\mathbf{x}}).
\end{eqnarray}
It is defined by the orthogonality relation
\begin{eqnarray}\label{U12}
\int d^{D}x' ~W_{ab}(t;{\mathbf{x}},{\mathbf{x}}')
W^{bc}(t;{\mathbf{x}}',{\mathbf{x}}'')=\delta^{D}({\mathbf{x}}-{\mathbf{x}}'')\delta^{\
c}_{a}.
\end{eqnarray}
As for the fermionic part of $\Omega(t)$, it  is defined similarly
by
\begin{eqnarray}\label{U13}
\Omega_{f}(t)\equiv\frac{1}{2}\int d^{D}x\ d^{D}x'\left(
W_{\bar{\psi}{\psi}}^{\alpha\beta}(t;{\mathbf{x}},{\mathbf{x}}')d\bar{\psi}_{\alpha}(t;{\mathbf{x}})\wedge
d\psi_{\beta}(t;{\mathbf{x}}')+W_{\psi\bar{\psi}}^{\alpha\beta}(t;{\mathbf{x}},{\mathbf{x}}')d\psi_{\beta}(t;{\mathbf{x}})
\wedge
d\bar{\psi}_{\alpha}(t;{\mathbf{x}}')\right).\nonumber\\
\end{eqnarray}
Here, $W_{\bar{\psi}\psi}$ and $W_{\psi\bar{\psi}}$ can be derived
using $X^{(m)}(t;\textbf{x})$, and $Y^{(m)}(t;\textbf{x})$ appearing
in (\ref{U8}). They include derivatives acting on $\bar{\psi}$ and
$\psi$, respectively. Using further
\begin{eqnarray}\label{U14}
W^{\alpha\beta}_{\psi\bar{\psi}}(t;{\mathbf{x}},{\mathbf{x}}')=W^{\alpha\beta}_{\bar{\psi}\psi}(t;{\mathbf{x}}',{\mathbf{x}}),
\end{eqnarray}
and $d\psi_{\beta}(t;\textbf{x}')\wedge
d\bar{\psi}_{\alpha}(t;\textbf{x})=d\bar\psi_{\alpha}(t;\textbf{x})\wedge
d{\psi}_{\beta}(t;\textbf{x}')$, the two-form (\ref{U13}) can be
brought in a simpler form
\begin{eqnarray}\label{U15}
\Omega_{f}(t) =\int d^{D}x\ d^{D}x'
W_{\bar{\psi}{\psi}}^{\alpha\beta}(t;{\mathbf{x}},{\mathbf{x}}')\
d\bar\psi_{\alpha}(t;{\mathbf{x}})\wedge
d{\psi}_{\beta}(t;{\mathbf{x}}').
\end{eqnarray}
The operators $W_{\bar{\psi}\psi}$ and $W_{\psi\bar{\psi}}$ can be
determined using the definition $X^{(m)}$ and $Y^{(m)}$ from
(\ref{U8}). Their inverse operators are denoted by
$W^{\psi\bar{\psi}}$ and $W^{\bar{\psi}\psi}$, respectively. They
can be determined using the orthogonality relations
\begin{eqnarray}\label{U16}
\int d^{D}x'\
W_{\bar{\psi}{\psi}}^{\alpha\beta}(t;{\mathbf{x}},{\mathbf{x}}')
W^{\psi\bar{\psi}}_{\beta\rho}(t;{\mathbf{x}}',{\mathbf{x}}'')&=&\delta^{D}({\mathbf{x}}-{\mathbf{x}}'')
\delta^{\alpha}_{\ \rho},\nonumber\\
\int d^{D}x'\
W_{\psi\bar{\psi}}^{\alpha\beta}(t;{\mathbf{x}},{\mathbf{x}}')
W^{\bar{\psi}\psi}_{\beta\rho}(t;{\mathbf{x}}',{\mathbf{x}}'')
&=&\delta^{D}({\mathbf{x}}-{\mathbf{x}}'')\delta^{\alpha}_{\ \rho}.
\end{eqnarray}
Similar to bosonic Poisson brackets, the Poisson brackets between
the fermionic fields $\psi$ and $\bar{\psi}$ are defined as
\begin{eqnarray}\label{U17}
\{\psi(t;{\mathbf{x}}),\bar{\psi}(t;{\mathbf{x}}')\}_{\mbox{\tiny{PB}}}&=&W^{\psi\bar{\psi}}(t;{\mathbf{x}},{\mathbf{x}}'),\nonumber\\
\{\bar{\psi}(t;{\mathbf{x}}),{\psi}(t;{\mathbf{x}}')\}_{\mbox{\tiny{PB}}}&=&W^{\bar{\psi}{\psi}}(t;{\mathbf{x}},{\mathbf{x}}').
\end{eqnarray}
The inverse operators $W^{\psi\bar{\psi}}$ and $W^{\bar{\psi}\psi}$
have, similar to their inverse operators $W_{\bar{\psi}\psi}$ and
$W_{\psi\bar{\psi}}$, the property
\begin{eqnarray}\label{U18}
W_{\alpha\beta}^{\psi\bar{\psi}}(t;{\mathbf{x}},{\mathbf{x}}')=W_{\alpha\beta}^{\bar{\psi}\psi}(t;{\mathbf{x}}',{\mathbf{x}}).
\end{eqnarray}
\subsection{Modified Poisson Brackets}
\par\noindent
As we have mentioned at the beginning of this section, in the
perturbative quantization introduced in \cite{ho-1, ho-2, eliezer},
the field variables $\psi,\bar{\psi}$ and $A_{\mu}$ are to be
redefined so that the corresponding Poisson brackets are the same as
in the ordinary QED. To do this, we will consider first the equation
of motion of free fermionic and bosonic fields arising from
(\ref{U2}),\footnote{Here, the gauge fixing parameter is chosen to
be $\xi=1$.}
\begin{eqnarray}\label{U19}
\gamma^{\mu}\partial_{\mu}\psi=0,\qquad
\partial_{\mu}\bar{\psi}\gamma^{\mu}=0,\qquad \Box A_{\mu}=0.
\end{eqnarray}
Then, using (\ref{U19}), we will replace the time derivatives of
fermionic and bosonic fields by the corresponding space derivatives
as
\begin{eqnarray}\label{U20}
\partial^{n}_{0}\psi_{\alpha}&=&\left\{
\begin{array}{rclcrcl}
n&=&2p,&\qquad&\partial_{0}^{2p}\psi_{\alpha}&=&\partial_{i}^{2p}\psi_{\alpha},\\
n&=&2p+1,&\qquad&\partial_{0}^{2p+1}\psi_{\alpha}&=&\left(-\gamma^{0}\gamma^{i}\right)_{\alpha\beta}\partial_{i}^{2p+1}\psi^{\beta},
\end{array}
\right.\nonumber\\
\partial^{n}_{0}\bar{\psi}_{\alpha}&=&\left\{
\begin{array}{rclcrcl}
n&=&2p,&\qquad&\partial_{0}^{2p}\bar{\psi}_{\alpha}&=&\partial_{i}^{2p}\bar{\psi}_{\alpha},\\
n&=&2p+1,&\qquad&\partial_{0}^{2p+1}\bar{\psi}_{\alpha}&=&\partial_{i}^{2p+1}\bar{\psi}^{\beta}\left(\gamma^{0}\gamma^{i}\right)_{\beta\alpha},
\end{array}
\right.\nonumber\\
\partial^{n}_{0}A^{\nu}&=&\left\{
\begin{array}{rclcrcl}
n&=&2p,&\qquad&\partial_{0}^{2p}A^{\mu}&=&\partial_{i}^{2p}A^{\mu},\\
n&=&2p+1,&\qquad&\partial_{0}^{2p+1}A^{\mu}&=&\partial_{i}^{2p}\dot
A^{\mu},
\end{array}
\right.,
\end{eqnarray}
where a summation over $i=1,\cdots, D$ is to be performed. The above
replacement lead to a modification of the Poisson brackets defined
in the previous section and eventually to a perturbative
quantization of the theory up to second order in the coupling
constant $e$.
\par
To modify the Poisson brackets of the fermionic and bosonic fields,
let us now consider $\delta S$ from (\ref{U5}) and replace the time
derivatives with the corresponding space derivatives using
(\ref{U20}). We arrive after a lengthy but straightforward
computation at
\begin{eqnarray}\label{U21}
\delta S&=&\sum_{p=0}^{\infty}\int d^{D}x
\left\{\left(\partial_{i}^{2p}\Hat{\Pi}_{\psi^{(2p)}}
+\partial_{i}^{2p+1}\Hat{\Pi}_{\psi^{(2p+1)}}
\gamma^{0}\gamma^{i}\right) \delta
\psi-\delta\bar{\psi}\left(\partial_{i}^{2p}\Hat{\Pi}_{\bar{\psi}^{(2p)}}^{\alpha}-\gamma^{0}\gamma^{i}\partial_{i}^{2p+1}\Hat{\Pi}_{\bar{\psi}^{(2p+1)}}\right)\right.\nonumber\\
&&\left.+\partial_{i}^{2p}\Hat{\Pi}_{A^{(2p)}}^{\mu}\delta
A_{\mu}+\partial_{i}^{2p}\Hat{\Pi}_{A^{(2p+1)}}^{\mu}\delta \dot
A_{\mu}\right\},
\end{eqnarray}
where the superscripts $(p)$ denote the $p$-th order space
derivative of the corresponding fields. Using (\ref{U21}) the
modified symplectic two-form is given by
\begin{eqnarray}\label{U22}
\hat{\Omega}(t)=\int d^{D}x\left(dP_{\psi}^{\alpha}\wedge
d\psi_{\alpha}-d \bar{\psi}_{\alpha}\wedge
dP_{\bar{\psi}}^{\alpha}+dP^{\mu}_{A}\wedge dA_{\mu}+dP^{\mu}_{\dot
A}\wedge d\dot A_{\mu}\right),
\end{eqnarray}
where comparing to (\ref{U21}), the modified momenta
$P_{\psi},P_{\bar{\psi}},P_{A}^\mu$ and $P_{\dot{A}}^\mu$ read
\begin{eqnarray*}
P_{\psi}&=&\sum\limits_{p=0}^{\infty}\left(\partial_{i}^{2p}\Hat{\Pi}_{\psi^{(2p)}}
+\partial_{i}^{2p+1}\Hat{\Pi}_{\psi^{(2p+1)}}\gamma^{0}\gamma^{i}\right),\nonumber\\
P_{\bar{\psi}}&=&\sum\limits_{p=0}^{\infty}\left(\partial_{i}^{2p}\Hat{\Pi}_{\bar{\psi}^{(2p)}}-
\gamma^{0}\gamma^{i}\partial_{i}^{2p+1}\Hat{\Pi}_{\bar{\psi}^{(2p+1)}}\right),\nonumber\\
\end{eqnarray*}
\begin{eqnarray}\label{U23}
P^{\mu}_{A}&=&\sum\limits_{p=0}^{\infty}\partial_{i}^{2p}\Hat{\Pi}_{A^{(2p)}}^{\mu},\nonumber\\
P^{\mu}_{\dot
A}&=&\sum\limits_{p=0}^{\infty}\partial_{i}^{2p}\Hat{\Pi}_{A^{(2p+1)}}^{\mu}.
\end{eqnarray}
Here, $\Hat{\Pi}_{\psi},
\Hat{\Pi}_{\bar{\psi}},\Hat{\Pi}_{A}^{\mu}$, and
$\Hat{\Pi}_{\dot{A}}^{\mu}$ are defined in (\ref{U6}). In a
perturbative quantization up to ${\cal{O}}(e^{2})$, the modified
momenta from (\ref{U23}) can be separated in a free and an
interaction part proportional to the coupling constant $e$,
\begin{eqnarray}\label{U24}
P_{\psi}=i\bar\psi\gamma^{0}+ie~\bar\psi \overleftarrow{{\cal
J}_{\psi}}\gamma^{0},&\qquad\qquad&
P_{\bar\psi}=-ie\gamma^{0}\overrightarrow{{\cal J}_{\bar\psi}}\psi,\nonumber\\
P^{\mu}_{A}=-\dot A^{\mu}-e\xi^{\mu}_{A},&\qquad\qquad&
P^{\mu}_{\dot A}=-e\xi^{\mu}_{\dot A},
\end{eqnarray}
where ${\cal{J}}_{\psi}, {\cal{J}}_{\bar{\psi}}, \xi_{A}^{\sigma}$,
and $\xi_{\dot{A}}^{\sigma}$ can be determined from
(\ref{U23}).\footnote{The notation in (\ref{U24}) is particularly
suitable for a Lagrangian of the form (\ref{U1})-(\ref{U2}).} In
(\ref{U24}), the operators ${\cal{J}}_{\psi}$ and
${\cal{J}}_{\bar{\psi}}$ include derivatives that act left on
$\bar{\psi}$ and right on $\psi$, respectively. This is denoted by
left and right arrows. Using (\ref{U24}), the modified symplectic
two-form $\hat{\Omega}(t)$ of modified QED described by the
Lagrangian (\ref{U1})-(\ref{U2}) is then given by
\begin{eqnarray}\label{U25}
\hat{\Omega}(t)&=&\int
d^{D}x\bigg\{d\bar\psi\wedge\bigg[i\gamma^{0}+ie\left(
\overleftarrow{ {\cal
J}_{\psi}}\gamma^{0}+\gamma^{0}\overrightarrow{{\cal
J}_{\bar\psi}}\right)\bigg]d\psi+ie~\bar\psi\left(\frac{\delta\overleftarrow{
{\cal J}_{\psi}}}{\delta A^{\rho}}\gamma^{0}\right)dA^{\rho}\wedge
d\psi\nonumber\\&&+ie~\bar\psi\left(\frac{\delta\overleftarrow{{\cal
J}_{\psi}}}{\delta \dot A^{\rho}}\gamma^{0}\right)d\dot
A^{\rho}\wedge d\psi+ie~d\bar\psi\wedge
dA^{\rho}\left(\gamma^{0}\frac{\delta\overrightarrow{{\cal
J}_{\bar\psi}}}{\delta A^{\rho}}\right)\psi+ie~d\bar\psi\wedge d\dot
A^{\rho}\left(\gamma^{0}\frac{\delta\overrightarrow{{\cal
J}_{\bar\psi}}}{\delta \dot A^{\rho}}\right)\psi\nonumber\\&&
+e\left(\frac{\delta\xi_{\mu}^{A}}{\delta A^{\nu}}
\right)dA^{\mu}\wedge dA^{\nu}+e\left(\frac{\delta\xi_{\mu}^{\dot
A}}{\delta\dot A^{\nu}}\right)d\dot A^{\mu}\wedge d\dot
A^{\nu}+\bigg[g_{\mu\nu}+e\left(\frac{\delta\xi_{\mu}^{A}}{\delta\dot
A^{\nu}}-\frac{\delta\xi_{\nu}^{\dot A}}{\delta
A^{\mu}}\right)\bigg]dA^{\mu}\wedge d\dot A^{\nu}\bigg\}\nonumber\\
&&+{\cal{O}}(e^{2}).
\end{eqnarray}
To determine the symplectic two form (\ref{U25}) we have used the
fact that for a theory described by (\ref{U1}) and (\ref{U2}),
$\xi_{A}$ and $\xi_{\dot{A}}$ are functions of $A_{\mu}$ and its
derivatives. To determine the corresponding Poisson algebra, let us
consider the modified symplectic two-form (\ref{U25}), which can be
formally given as
\begin{eqnarray}\label{U28}
\hat{\Omega}(t)&=&\int d^{D}x
\left(\hat{W}^{\bar{\psi}\psi}_{\alpha\beta}\
d\bar{\psi}^{\alpha}\wedge d\psi^{\beta}
+\hat{W}^{A\psi}_{\mu\alpha}\ dA^{\mu}\wedge
d\psi^{\alpha}+\hat{W}^{\dot{A}\psi}_{\mu\alpha}\
d\dot{A}^{\mu}\wedge
d\psi^{\alpha}+\hat{W}^{\bar{\psi}A}_{\alpha\mu}\
d\bar{\psi}^{\alpha}\wedge
dA^{\mu}\right.\nonumber\\
&&\left.+\hat{W}^{\bar{\psi}\dot{A}}_{\alpha\mu}\
d\bar{\psi}^{\alpha}\wedge
d\dot{A}^{\mu}+\hat{W}^{\dot{A}A}_{\mu\nu}\ d\dot{A}^{\mu}\wedge
dA^{\nu}+\hat{W}^{AA}_{\mu\nu}\ dA^{\mu}\wedge
dA^{\nu}+\hat{W}^{\dot{A}\dot{A}}_{\mu\nu}\ d\dot{A}^{\mu}\wedge
d\dot{A}^{\nu}\right)+{\cal{O}}(e^{2}).\nonumber\\
\end{eqnarray}
Here, the coefficients $\hat{W}_{ab}$ with $a,b\in\{\bar{\psi},\psi,
A,\dot{A}\},$ can be read from (\ref{U25}). Note that in
(\ref{U28}), $\mu,\nu=0,\cdots,D$ are the space-time indices and
$\alpha, \beta=1,\cdots,2^{\frac{D+1}{2}}$ are the spinor indices.
Comparing with the original symplectic two-form $\Omega(t)$ from
(\ref{U9}), whose bosonic and fermionic parts are given in
(\ref{U10}) and (\ref{U13}), new bases appear in the
(perturbatively) modified phase space of the theory.  Note that
among the terms in $\hat{\Omega}(t)$ from (\ref{U25}) and
(\ref{U28}), the terms proportional to $dA^{\mu}\wedge dA^{\nu}$ and
$d\dot{A}^{\mu}\wedge d\dot{A}^{\nu}$ survive only if their
coefficients are antisymmetric in $\mu$ and $\nu$. According to our
definitions from previous section, the modified Poisson brackets of
the fermionic and bosonic degrees of freedom can be determined by
inverting $\hat{W}_{ab}$ using the orthonormality relations similar
to (\ref{U12}) and (\ref{U16}). Up to ${\cal{O}}(e^{2})$, the
modified Poisson brackets read
\begin{eqnarray}\label{U29}
\hat{W}_{\mu\nu}^{\dot A A}(t;\textbf{x},\textbf{x}')&=&\{\dot
A_{\mu}(t;\textbf{x}),A_{\nu}(t;\textbf{x}')\}_{\mbox{\tiny{PB}}}=\bigg[g_{\mu\nu}+e\left(\frac{\delta\xi^{\dot
A }_{\nu}}{\delta A^{\mu}}-\frac{\delta\xi^{A}_{\mu}}{\delta\dot
A^{\nu}}\right)\bigg]\delta^{D}(\textbf{x}-\textbf{x}'),\nonumber\\
\hat{W}_{\mu\nu}^{A\dot A}(t;\textbf{x},\textbf{x}')&=&\{
A_{\mu}(t;\textbf{x}),\dot
A_{\nu}(t;\textbf{x}')\}_{\mbox{\tiny{PB}}}=\bigg[-g_{\mu\nu}+e\left(\frac{\delta\xi^{
A }_{\nu}}{\delta \dot A^{\mu}}-\frac{\delta\xi^{\dot
A}_{\mu}}{\delta
A^{\nu}}\right)\bigg]\delta^{D}(\textbf{x}-\textbf{x}'),\nonumber\\
\hat{W}_{\mu\nu}^{\dot A\dot A}(t;\textbf{x},\textbf{x}')&=&\{\dot
A_{\mu}(t;\textbf{x}),\dot
A_{\nu}(t;\textbf{x}')\}_{\mbox{\tiny{PB}}}=e\left(\frac{\delta\xi^{A}_{\mu}}{\delta
A^{\nu}}-\frac{\delta\xi^{ A}_{\nu}}{\delta
A^{\mu}}\right)\delta^{D}(\textbf{x}-\textbf{x}'),\nonumber\\
\hat{W}_{\mu\nu}^{AA}(t;\textbf{x},\textbf{x}')&=&\{
A_{\mu}(t;\textbf{x}),
A_{\nu}(t;\textbf{x}')\}_{\mbox{\tiny{PB}}}=e\left(\frac{\delta\xi^{\dot
A}_{\mu}}{\delta\dot A^{\nu}}-\frac{\delta\xi^{\dot A}_{\nu}}{\delta
\dot A^{\mu}}\right)\delta^{D}(\textbf{x}-\textbf{x}'),\nonumber\\
\hat{W}_{\mu\alpha}^{A\bar\psi}(t;\textbf{x},\textbf{x}')&=&\{
A_{\mu}(t;\textbf{x}),\bar\psi_{\alpha}(t;\textbf{x}')\}_{\mbox{\tiny{PB}}}=e\bar\psi^{\beta}\left(\frac{\delta\overleftarrow{{\cal
J}}_{\psi}}{\delta \dot
A^{\mu}}\right)_{\beta\alpha}\delta^{D}(\textbf{x}-\textbf{x}'),\nonumber\\
\hat{W}_{\mu\alpha}^{A\psi}(t;\textbf{x},\textbf{x}')&=&\{A_{\mu}(t;\textbf{x}),\psi_{\alpha}(t;\textbf{x}')\}_{\mbox{\tiny{PB}}}=e\left(\frac{\delta\overrightarrow{{\cal
J}}_{\bar\psi}}{\delta\dot
A^{\mu}}\right)_{\alpha\beta}\psi^{\beta}\delta^{D}(\textbf{x}-\textbf{x}'),\nonumber\\
\hat{W}_{\mu\alpha}^{\dot
A\bar\psi}(t;\textbf{x},\textbf{x}')&=&\{\dot
A_{\mu}(t;\textbf{x}),\bar\psi_{\alpha}(t;\textbf{x}')\}_{\mbox{\tiny{PB}}}=-e\bar\psi^{\beta}\left(\frac{\delta\overleftarrow{{\cal
J}}_{\psi}}{\delta
A^{\mu}}\right)_{\beta\alpha}\delta^{D}(\textbf{x}-\textbf{x}'),\nonumber\\
\hat{W}_{\mu\alpha}^{\dot A\psi}(t;\textbf{x},\textbf{x}')&=&\{\dot
A_{\mu}(t;\textbf{x}),\psi_{\alpha}(t;\textbf{x}')\}_{\mbox{\tiny{PB}}}=-e\left(\frac{\delta\overrightarrow{{\cal
J}}_{\bar\psi}}{\delta
A^{\mu}}\right)_{\alpha\beta}\psi^{\beta}\delta^{D}(\textbf{x}-\textbf{x}'),\nonumber\\
\hat{W}_{\alpha\beta}^{\psi\bar\psi}(t;\textbf{x},\textbf{x}')&=&\{\psi_{\alpha}(t;\textbf{x}),\bar\psi_{\beta}(t;\textbf{x}')\}_{\mbox{\tiny{PB}}}=\bigg[-i\gamma^{0}+ie\left(\overrightarrow{{\cal
J}_{\bar\psi}}\gamma^{0}+\gamma^{0}\overleftarrow{{\cal
J}_{\psi}}\right)\bigg]_{\alpha\beta}\delta^{D}(\textbf{x}-\textbf{x}').
\end{eqnarray}
\subsection{Field Redefinition and Dirac Quantization}
\par\noindent
In this section, we will redefine the fermionic and bosonic field
variables so that their corresponding Poisson brackets are the same
as in ordinary QED. To do this let us first define the modified
gauge field $\tilde{A}_{\mu}$ and its corresponding modified
canonical momentum $\tilde{\Pi}_{\mu}$ as
\begin{eqnarray}\label{U30}
\tilde{A}_{\mu}\equiv A_{\mu}-e\xi^{\dot A}_{\mu},&\qquad&
\tilde{\Pi}_{\mu}\equiv - \dot A_{\mu}-e\xi_{\mu}^{A},
\end{eqnarray}
and determine the bosonic Poisson bracket
\begin{eqnarray*}
\{\tilde{A}_{\mu}(t;{\mathbf{x}}),\tilde{\Pi}_{\nu}(t;{\mathbf{x}}')\}_{\mbox{\tiny{PB}}},
\end{eqnarray*}
using the Poisson brackets (\ref{U29}). In (\ref{U30}),
$\xi^{A}_{\mu}$ and $\xi^{\dot{A}}_{\mu}$ are the same as introduced
in (\ref{U24}). Replacing $\tilde{\Pi}_{\nu}$ from (\ref{U30}) in
the above Poisson bracket, we arrive first at
\begin{eqnarray}\label{U31}
\{\tilde{A}_{\mu}(t;{\mathbf{x}}),\tilde{\Pi}_{\nu}(t;{\mathbf{x}}')\}&=&-\{A_{\mu}(t;{\mathbf{x}}),
\dot
A_{\nu}(t;{\mathbf{x}}')\}-e\{A_{\mu}(t;{\mathbf{x}}),\xi^{A}_{\nu}(t;{\mathbf{x}}')\}+e\{\xi_{\mu}^{\dot
A}(t;{\mathbf{x}}),\dot A_{\nu}(t;{\mathbf{x}}')\}+{\cal
O}(e^{2}),\nonumber\\
\end{eqnarray}
where we have skipped the subscript PB (Poisson Bracket). To
evaluate the first two terms on the r.h.s. (right hand side) of the
above relations, we use the Poisson brackets (\ref{U29}). The
remaining three terms in (\ref{U31}) can be determined using the
standard definition of the Poisson bracket of two functionals
${\cal{F}}, {\cal{G}}$ that depend on the dynamical variables
$(\eta,\pi_{\eta})$ including the fermionic and bosonic field
variables and their derivatives \cite{reyes-2},
\begin{eqnarray}\label{U32}
\{{\cal{F}}(t;{\mathbf{x}}),{\cal{G}}(t;{\mathbf{x}}')\}_{\mbox{\tiny{PB}}}&\equiv&\int
d^{D}z d^{D}z'\left(\frac{\delta{\cal{F}}(t;{\mathbf{x}})}{\delta
\eta(t;{\mathbf{z}})}\frac{\delta{\cal{G}}(t;{\mathbf{x}}')}{\delta\pi_{\eta}(t;{\mathbf{z}}')}\mp\frac{\delta{\cal{F}}(t;{\mathbf{x}})}{\delta
\pi_{\eta}(t;{\mathbf{z}}')}\frac{\delta{\cal{G}}(t;{\mathbf{x}}')}{\delta\eta(t;{\mathbf{z}})}\right)\{\eta(t;{\mathbf{z}}),
\pi_{\eta}(t;{\mathbf{z}}')\}_{\mbox{\tiny{PB}}}.\nonumber\\
\end{eqnarray}
Here, the minus (plus) sign corresponds to bosonic (fermionic)
fields $\eta$.  Separating the space and time components of the
indices $\mu$ and $\nu$ in (\ref{U31}) and after a lengthy but
straightforward calculation, the canonical equal-time Poisson
bracket of $\tilde{A}_{\mu}$ and its conjugate momentum
$\tilde{\Pi}_{\nu}$ reads
\begin{eqnarray}\label{U33}
\{\tilde{A}_{\mu}(t;{\mathbf{x}}),\tilde{\Pi}_{\nu}(t;{\mathbf{x}}')\}_{\mbox{\tiny{PB}}}=g_{\mu\nu}\delta^{D}({\mathbf{x}}-{\mathbf{x}}')+{\cal{O}}(e^{2}).
\end{eqnarray}
This is, up to ${\cal{O}}(e^{2})$, the standard Poisson bracket of
ordinary QED which can be quantized in the standard Dirac
quantization procedure, i.e. replacing
\begin{eqnarray}\label{U34}
\begin{array}{rclcl}
\{\ \cdot\ ,\ \cdot\ \}_{\mbox{\tiny{PB}}}&\to&-i[\ \cdot\ ,\ \cdot\
]_{\mbox{\tiny{DB}}},&\qquad &\mbox{for bosons,}\\
\{\ \cdot\ ,\ \cdot\ \}_{\mbox{\tiny{PB}}}&\to&-i\{\ \cdot\ ,\
\cdot\ \}_{\mbox{\tiny{DB}}},&\qquad& \mbox{for fermions},
\end{array}
\end{eqnarray}
we arrive at
\begin{eqnarray}\label{U35}
[\tilde{A}_{\mu}(t;{\mathbf{x}}),\tilde{\Pi}_{\nu}(t;{\mathbf{x}}')]_{\mbox{\tiny{DB}}}=ig_{\mu\nu}\delta^{D}({\mathbf{x}}-{\mathbf{x}}')+{\cal{O}}(e^{2}).
\end{eqnarray}
Here, the subscript DB is the Dirac bracket. Similarly, one finds
\begin{eqnarray}
[\tilde{A}_{\mu}(t;{\mathbf{x}}),\tilde{A}_{\nu}(t;{\mathbf{x}}')]_{\mbox{\tiny{DB}}}=[\tilde{\Pi}_{\mu}(t;{\mathbf{x}}),\tilde{\Pi}_{\nu}(t;{\mathbf{x}}')]_{\mbox{\tiny{DB}}}={\cal{O}}(e^{2}).
\end{eqnarray}
What concerns the redefinition of the fermionic fields, let us
define $\psi$ and $\bar{\psi}$ as
\begin{eqnarray}\label{U36}
\tilde{\psi}=\psi+e\overrightarrow{{\cal
J}_{\bar\psi}}\psi,\qquad\tilde{\bar\psi}=\bar\psi+e\bar\psi\overleftarrow{{\cal
J}_{\psi}},
\end{eqnarray}
where ${\cal{J}}_{\bar\psi}$ and ${\cal{J}}_{\psi}$ are the same as
introduced in (\ref{U24}). Replacing (\ref{U36}) in the
corresponding Poisson bracket
$$
\{\tilde{\psi}(t;{\mathbf{x}}),\tilde{\bar\psi}(t;{\mathbf{x}}'\}_{\mbox{\tiny{PB}}},
$$
and using the modified algebra from (\ref{U29}) and the definition
of Poisson bracket from (\ref{U32}), we arrive at
\begin{eqnarray}\label{U37}
\{\tilde{\psi}(t;{\mathbf{x}}),\tilde{\bar\psi}(t;{\mathbf{x}}')\}_{\mbox{\tiny{PB}}}&=&
\{\psi(t;{\mathbf{x}}),\bar\psi(t;{\mathbf{x}}')\}+e\{\psi(t;{\mathbf{x}}),\bar\psi(t;{\mathbf{x}}')\}\overleftarrow{{\cal
J}_{\psi}}+e\overrightarrow{{\cal
J}_{\bar\psi}}\{\psi(t;{\mathbf{x}}),\bar\psi(t;{\mathbf{x}}')\}+{\cal
O}(e^{2})\nonumber\\
&=&-i\gamma_{0}\delta^{D}({\mathbf{x}}-{\mathbf{x}}')+{\cal{O}}(e^{2}).
\end{eqnarray}
This leads after replacing the Poisson bracket by the Dirac bracket
using (\ref{U34}) to
\begin{eqnarray}\label{U38}
\{\tilde{\bar\psi}(t;{\mathbf{x}}),\tilde{\psi}(t;{\mathbf{x}}')\}_{\mbox{\tiny{DB}}}=\gamma^{0}\delta^{D}({\mathbf{x}}-{\mathbf{x}}')+{\cal{O}}(e^{2}),
\end{eqnarray}
which is up to ${\cal{O}}(e^{2})$ the ordinary canonical equal-time
anticommutation relation of ordinary QED. Apart from (\ref{U37}), it
can easily be checked that the Poisson brackets of the modified
gauge fields $\tilde{A}_{\mu}$ from (\ref{U30}) with the modified
fermions $\tilde{\psi},\tilde{\bar{\psi}}$ from (\ref{U36}), i.e.
$\{\tilde{A}_{\mu}(t;{\mathbf{x}}),\tilde{\psi}(t;{\mathbf{x}}')\}$
and
$\{\tilde{A}_{\mu}(t;{\mathbf{x}}),\tilde{\bar{\psi}}(t;{\mathbf{x}}')\}$,
vanish, as is expected also from ordinary QED.
\par
In the next section, we will first introduce the Lagrangian density
of two-dimensional space-time noncommutative QED, as an example of a
modified QED including higher order time derivatives. The modified
Poisson brackets of fermionic and bosonic degrees of freedom will be
then determined perturbatively up to order two in the coupling
constant $e$ and order three in the noncommutativity parameter
$\theta$.
\section{Modified Poisson brackets of two-dimensional space-time
noncommutative QED}
\setcounter{equation}{0}
\par\noindent
The noncommutative gauge theory is characterized by the replacement
of the familiar product of functions with the $\star$-product
defined by
\begin{eqnarray}\label{P1}
f(x)\star g(x)\equiv
\exp\left(\frac{i\theta_{\mu\nu}}{2}\frac{\partial}{\partial\xi_{\mu}}\frac{\partial}{\partial
\zeta_{\nu}}\right)\ f(x+\xi)g(x+\zeta)\Bigg|_{\xi=\zeta=0}.
\end{eqnarray}
In two space-time dimensions, $\theta_{\mu\nu}$, being an
antisymmetric matrix and reflecting the noncommutativity of space
and time coordinates, reduces to
$\theta_{\mu\nu}=\theta\epsilon_{\mu\nu}$, where $\epsilon_{\mu\nu}$
is the two-dimensional Levi-Civita symbol. The $\star$-product
satisfies the identity
\begin{eqnarray}\label{P2}
\int_{-\infty}^{+\infty}d^{d}x\ f(x)\star
g(x)=\int_{-\infty}^{+\infty}d^{d}x\ g(x)\star
f(x)=\int_{-\infty}^{+\infty}d^{d}x\ f(x)g(x),
\end{eqnarray}
and is associative
\begin{eqnarray}\label{P3}
\int_{-\infty}^{+\infty}d^{d}x\ (f\star g\star
h)(x)=\int_{-\infty}^{+\infty}d^{d}x\ (h\star f\star
g)(x)=\int_{-\infty}^{+\infty}d^{d}x\ (g\star h\star f)(x).
\end{eqnarray}
Here, $d$ is the dimension of space-time coordinates.  Let us
consider the Lagrangian density of two-dimensional space-time
noncommutative QED including the fermionic and the bosonic fields
\begin{eqnarray}\label{P4}
{\cal{L}}=i\bar{\psi}\star\gamma^{\mu}\partial_{\mu} \psi-e\
\bar{\psi}\star \gamma^{\mu}A_{\mu}\star
\psi-\frac{1}{4}F_{\mu\nu}\star
F^{\mu\nu}-\frac{1}{2\xi}(\partial_{\mu} A^{\mu})\star
(\partial_{\nu} A^{\nu}),
\end{eqnarray}
where $F_{\mu\nu}=\partial_{\mu} A_{\nu}-\partial_{\nu} A_{\mu}+ie\
[A_{\mu},A_{\nu}]_{\star}$, and
$[A_{\mu},A_{\nu}]_{\star}=A_{\mu}\star A_{\nu}-A_{\nu}\star
A_{\mu}$. The corresponding Euler-Lagrangian equation of motion for
the fermionic and bosonic fields are given by
\begin{eqnarray}\label{P5}
\gamma^{\mu}\partial_{\mu}\psi+ie\gamma^{\mu}A_{\mu}\star\psi&=&0,\nonumber\\
\partial_{\mu}\bar{\psi}\gamma^{\mu}-ie\bar{\psi}\gamma^{\mu}\star
A_{\mu}&=&0,\nonumber\\
D_{\mu}F^{\mu\nu}+\frac{1}{\xi}\partial^{\nu}\partial_{\mu}A^{\mu}&=&eJ^{\nu},
\end{eqnarray}
with $D_{\mu}=\partial_{\mu}+ie[A_{\mu},\cdot]_{\star}$. The
covariant $U_{V}(1)$ vector current $J^{\nu}$ in (\ref{P5}) is
defined by
\begin{eqnarray}\label{P6}
J^{\mu}(x)\equiv -\psi_{\beta}(x)\star\bar\psi_{\alpha}(x)
(\gamma^{\mu})^{\alpha\beta}.
\end{eqnarray}
In two space-time dimensions, the above theory can be regarded as a
higher order time derivative theory, where the higher order time
derivatives appear only in the interaction part. To show this, let
us separate the Lagrangian (\ref{P4}) in a free and an interaction
part ${\cal{L}}={\cal{L}}_{0}+{\cal{L}}_{\mbox{\tiny{int}}}$. Using
the relations (\ref{P2}) and (\ref{P3}), the $\star$-product in the
free part of the Lagrangian (\ref{P4}) can be removed, leaving us
with the ordinary free part of commutative QED Lagrangian
\begin{eqnarray}\label{P7}
{\cal{L}}_{0}=\bar{\psi}i\gamma^{\mu}\partial_{\mu}\psi-\frac{1}{4}
{\cal{F}}_{\mu\nu}{\cal{F}}^{\mu\nu}-\frac{1}{2\xi}(\partial_{\mu}
A^{\mu})^{2},
\end{eqnarray}
where ${\cal{F}}_{\mu\nu}$ is the commutative field strength tensor
[see below (\ref{U2}) for its definition]. The interaction part of
(\ref{P4}), after removing one of the $\star$-products, can be
separated into two parts, ${\cal{L}}_{\mbox{\tiny{int}}}\equiv
{\cal{L}}^{(1)}_{\mbox{\tiny{int}}}+{\cal{L}}_{\mbox{\tiny{int}}}^{(2)}$.
The first part includes the interaction of fermions $\psi$ and
$\bar{\psi}$ with the gauge field $A_{\mu}$. It is given by
\begin{eqnarray}\label{P8}
{\cal{L}}_{\mbox{\tiny{int}}}^{(1)}&=&+e\psi_{\beta}\star\bar{\psi}_{\alpha}(\gamma^{\lambda})_{\alpha\beta}A_{\lambda},\nonumber\\
&=&+e\sum\limits_{n=0}^{\infty}\left(\frac{i\theta}{2}\right)^{n}\frac{1}{n!}\epsilon^{\mu_{1}\nu_{1}}\cdots
\epsilon^{\mu_{n}\nu_{n}}\left(\partial_{\mu_{1}}\cdots\partial_{\mu_{n}}\psi_{\beta}\right)\left(\partial_{\nu_{1}}\cdots\partial_{\nu_{n}}\bar{\psi}_{\alpha}\right)
(\gamma^{\lambda})^{\alpha\beta}A_{\lambda}.
\end{eqnarray}
The second part consists of bosons self-interaction from the gauge
kinetic term in (\ref{P4}). It is given by
\begin{eqnarray}\label{P9}
{\cal{L}}_{\mbox{\tiny{int}}}^{(2)}&=&-\frac{ie}{2}{\cal{F}}_{\mu\nu}\big[A^{\mu},A^{\nu}\big]_{\star}+\frac{e^{2}}{4}\big[A_{\mu},A_{\nu}\big]_{\star}\big[A^{\mu},A^{\nu}\big]_{\star},\nonumber\\
&=&-ie{\cal{F}}_{\mu\nu}\sum\limits_{p=0}^{\infty}\left(\frac{i\theta}{2}\right)^{2p+1}\frac{1}{(2p+1)!}\epsilon^{\alpha_{1}\beta_{1}}\cdots\epsilon^{\alpha_{2p+1}\beta_{2p+1}}
\left(\partial_{\alpha_{1}}\cdots\partial_{\alpha_{2p+1}}A^{\mu}\right)(\partial_{\beta_{1}}\cdots\partial_{\beta_{2p+1}}A^{\nu})\nonumber\\
&+&e^{2}\sum\limits_{p,s=0}^{\infty}\left(\frac{i\theta}{2}\right)^{2p+2s+2}\frac{1}{(2p+1)!(2s+1)!}
\epsilon^{\rho_{1}\sigma_{1}}\cdots\epsilon^{\rho_{2p+1}\sigma_{2p+1}}\epsilon^{\alpha_{1}\beta_{1}}\cdots\epsilon^{\alpha_{2s+1}\beta_{2s+1}}\nonumber\\
&&\times
\left(\partial_{\rho_{1}}\cdots\partial_{\rho_{2p+1}}A_{\mu}\right)(\partial_{\sigma_{1}}\cdots\partial_{\sigma_{2p+1}}A_{\nu})
\left(\partial_{\alpha_{1}}\cdots\partial_{\alpha_{2s+1}}A^{\mu}\right)(\partial_{\beta_{1}}\cdots\partial_{\beta_{2s+1}}A^{\nu}).
\end{eqnarray}
In (\ref{P8}) and (\ref{P9}), the definition of the $\star$-product
from (\ref{P1}) is used. Using (\ref{U6}), the canonical conjugate
momenta corresponding to $\psi$, $\bar{\psi}$ and $A_{\mu}$ are
given by
\begin{eqnarray}\label{P10}
\Pi_{A^{(m)}}^{\sigma}&=&-\dot{A}^{\sigma}\delta^{m0}\nonumber\\
&& -ie
\sum_{k=0}^{\infty}\sum_{\ell=0}^{k}\sum\limits_{r=0}^{\ell}\sum\limits_{s=0}^{k-\ell}\left(\frac{i\theta}{2}\right)^{k+m+1}\frac{[(-1)^{\ell+1}+(-1)^{k+m+\ell+1}]}{(k+m+1)!}
\left(
\begin{array}{c}
\ell\\r
\end{array}
\right)\left(
\begin{array}{c}
k-\ell\\s
\end{array}
\right) \nonumber\\
&&\qquad\times\left(\partial_{0}^{r+k-\ell}\partial_{x}^{s+\ell+m+1}
A_{\mu}\right)\partial_{0}^{\ell-r}\partial_{x}^{k-\ell-s}{\cal{F}}^{\mu\sigma}\nonumber\\
&&-ie\sum_{p=0}^{\infty}\left(\frac{i\theta}{2}\right)^{2p+1}\frac{\epsilon^{\alpha_{1}\beta_{1}}\cdots\epsilon^{\alpha_{2p+1}\beta_{2p+1}}}{(2p+1)!}
\left(\partial_{\alpha_{1}}\cdots\partial_{\alpha_{2p+1}}A^{0}\right)
(\partial_{\beta_{1}}\cdots\partial_{\beta_{2p+1}}A^{\sigma})\delta^{m0}+{\cal O}(e^{2}),\nonumber\\
\Pi_{\psi^{(m)}}^{\beta}&=&-e
\sum\limits_{k=0}^{\infty}\sum\limits_{\ell=0}^{k}\sum\limits_{r=0}^{\ell}\sum\limits_{s=0}^{k-\ell}
\left(\frac{i\theta}{2}\right)^{m+k+1}\frac{(-1)^{\ell}}{(m+k+1)!}
\left(
\begin{array}{c}
\ell\\r
\end{array}
\right)\left(
\begin{array}{c}
k-\ell\\s
\end{array}
\right) \left(\partial_{x}^{s+m+\ell+1}\partial_{0}^{\
r+k-\ell}\bar{\psi}_{\alpha}\right)
\nonumber\\
&&\qquad\times \left(\partial_{0}^{\ell-r}
\partial_{x}^{k-\ell-s}A_{\lambda}\right)\left(\gamma^{\lambda}\right)^{\alpha\beta}+i\delta^{m0}\bar{\psi}_{\alpha}(\gamma_{0})^{\alpha\beta},
\nonumber\\
\Pi_{\bar{\psi}^{(m)}}^{\alpha}&=&-e
\sum\limits_{k=0}^{\infty}\sum\limits_{\ell=0}^{k}\sum\limits_{r=0}^{\ell}\sum\limits_{s=0}^{k-\ell}
\left(\frac{i\theta}{2}\right)^{m+k+1}\frac{(-1)^{k+m+\ell+1}}{(m+k+1)!}
\left(
\begin{array}{c}
\ell\\r
\end{array}
\right)\left(
\begin{array}{c}
k-\ell\\s
\end{array}
\right) \left(\partial_{x}^{s+m+\ell+1}\partial_{0}^{\
r+k-\ell}{\psi}_{\beta}\right)
\nonumber\\
&&\qquad\times \left(\partial_{0}^{\ell-r}
\partial_{x}^{k-\ell-s}A_{\lambda}\right)\left(\gamma^{\lambda}\right)^{\alpha\beta},
\end{eqnarray}
where in $\Pi_{A^{(m)}}^{\sigma}$ the gauge fixing parameter is
chosen to be $\xi=1$, and the terms of order $e^{2}$ are neglected.
In what follows, we will use the notations introduced in Sect. II to
determine separately the corresponding Poisson brackets to the gauge
field $A_{\mu}$ and the fermionic fields $\bar{\psi},\psi$ up to
order ${\cal{O}}(e^{2},\theta^{3})$.
\subsection{Poisson Brackets of Gauge Fields}
\par\noindent
Using (\ref{P9}), the Lagrangian density of the gauge fields is, up
to order ${\cal{O}}(e^{2},\theta^{3})$, given by
\begin{eqnarray}\label{P11}
{\cal
L}_{g}&=&-\frac{1}{4}{\cal{F}}_{\mu\nu}{\cal{F}}^{\mu\nu}-\frac{1}{2}(\partial_{\mu}
A^{\mu})^{2}+\frac{e\theta}{2}\epsilon^{\alpha\beta}\partial_{\alpha}A^{\mu}\partial_{\beta}A^{\nu}
{\cal{F}}_{\mu\nu}+{\cal{O}}(e^{2},\theta^{3}),\nonumber\\
&=&-\frac{1}{2}(\partial_{\mu}A_{\nu})(\partial^{\mu}A^{\nu})+e\theta\left((\partial_{x}A_{0})\dot
A_{1}-(\partial_{x}A_{1})\dot A_{0}\right){\cal
F}_{01}+{\cal{O}}(e^{2},\theta^{3}),
\end{eqnarray}
where an integration by part is performed in the
$\theta$-independent part of the Lagrangian. In this order of
perturbative expansion, the above Lagrangian includes only first
order time derivatives of $A_{\mu}$. It is therefore not necessary
to modify the theory in the sense of replacing the higher order time
derivatives by the corresponding space derivatives using the
corresponding equation of motion $\Box A_{\mu}=0$ from (\ref{U19})
and the resulting relations from (\ref{U20}).\footnote{According to
our notations from previous section, in this case the ``hatted'' and
``unhatted'' quantities are equal up to
${\cal{O}}(e^{2},\theta^{3})$.} In what follows, we will
nevertheless determine the symplectic two-forms and the
corresponding Poisson brackets using the formulation introduced in
Sect. II. As it turns out the Poisson brackets receives corrections
of order $e\theta$, that vanish by taking the commutative limit
$\theta\to 0$. To start, let us consider (\ref{U10}) and choose
$z^{a}=(A^{0}, A^{1},\dot{A}^{0},\dot{A}^{1})$ as the phase space
variables. The bosonic symplectic two form is given by
\begin{eqnarray}\label{P12}
\Omega_{g}(t)=\frac{1}{2}\int dx\ dx'\ W_{ab}(t;x,x')\
dz^{a}(t;x)\wedge dz^{b}(t;x'),
\end{eqnarray}
or equivalently by
$$
\Omega_{g}(t)=\int dx\ \left(Z^{(0)}+Z^{(1)}\right),
$$
where, according to (\ref{U8}), $Z^{(m)}$ is given as
$Z^{(m)}=d\Pi_{A^{(m)}}^{0}\wedge
dA_{0}^{(m)}+d\Pi_{A^{(m)}}^{1}\wedge dA_{1}^{(m)}$. Using the
general definition of $\Pi^{\sigma}_{A^{(m)}}$ from (\ref{P10}), the
momenta $\Pi_{\sigma}^{A^{(m)}}$ in the order
${\cal{O}}(e^{2},\theta^{3})$ read
\begin{eqnarray}\label{P13}
P^{A}_{0}=\Pi^{A}_{0}&=& -\dot{A}_{0}-e\theta
{\cal{F}}_{01}\left(\partial_{x}A_{1}\right)+{\cal{O}}(e^{2},\theta^{3}),\nonumber\\
P^{A}_{1}=\Pi^{A}_{1}&=&-\dot{A}_{1}-e\theta\left\{2\dot{A}_{1}
\partial_{x}A_{0}-(\partial_{x}A_{0})^{2}-
\dot{A}_{0}\partial_{x}A_{1}
\right\}+{\cal{O}}(e^{2},\theta^{3}),\nonumber\\
P^{\dot{A}}_{0}=\Pi^{\dot{A}}_{0}&=&{\cal{O}}(e,\theta^{3}),\nonumber\\
P^{\dot{A}}_{1}=\Pi^{\dot{A}}_{1}&=&{\cal{O}}(e,\theta^{3}).
\end{eqnarray}
Here, $\partial_{x}\equiv \frac{\partial}{\partial x^{1}}$ and
$A^{(m=0)}_{\mu}$ and $A^{(m=1)}_{\mu}$ are denoted by $A_{\mu}$ and
$\dot{A}_{\mu}$, respectively. Combining the above expressions we
arrive at
\begin{eqnarray}\label{P14}
Z^{(0)}&=& d \left(-\dot{A}_{0}+e\theta\left(
{\cal{F}}_{10}\partial_{x}A_{1} \right)\right)\wedge
dA_{0}+\left\{d\dot{A}_{1}+ e\theta\left(2\
d(\dot{A}_{1}\partial_{x}A_{0})-2(\partial_{x}A_{0})d(\partial_{x}A_{0})\right.\right.
\nonumber\\
&&\left.\left.-d(\dot
A_{0}\partial_{x}A_{1})\right)\right\}\wedge dA_{1}+{\cal{O}}(e^{2},\theta^{3}),\nonumber\\
Z^{(1)}&=&{\cal{O}}(e,\theta^{3}).
\end{eqnarray}
Performing appropriate integrations by part and using the
antisymmetry of bosonic wedge product, the expressions in
(\ref{P14}) can be simplified and we arrive up to
${\cal{O}}(e^{2},\theta^{3})$ at
\begin{eqnarray}\label{P15}
\lefteqn{W_{ab}(t;x,x')\simeq}\nonumber\\
&\simeq&\left(
  \begin{array}{ccccccc}
    0 && -e\theta\left({\cal{F}}_{01}\partial_{x}-\partial_{x}{\cal{F}}_{01}
\right) && 1 && -e \theta(\partial_{x}A_{1}) \\
    -e\theta\left({\cal{F}}_{01}\partial_{x}+2\partial_{x}{\cal{F}}_{01}\right) && 0
    &&
    -e\theta(\partial_{x}A_{1}) && -(1+2e\theta(\partial_{x}A_{0})) \\
    -1 && e\theta(\partial_{x}A_{1})&& 0 && 0 \\
    e\theta(\partial_{x}A_{1})&&
    1+2e\theta(\partial_{x}A_{0}) && 0 && 0 \\
  \end{array}
\right)\delta(x-x').\nonumber\\
\end{eqnarray}
To determine the bosonic Poisson brackets the above matrix is to be
inverted. After a lengthy but straightforward computation we arrive
at the inverse matrix $W_{ab}^{-1}(t;x,x')$ up to
${\cal{O}}(e^{2},\theta^{3})$
\begin{eqnarray}\label{P16}
\lefteqn{W_{ab}^{-1}(t;x,x')\simeq}\nonumber\\
&\simeq&\left(
  \begin{array}{ccccccc}
    0 && 0 && -1 && e \theta(\partial_{x}A_{1}) \\
    0 && 0 && e\theta(\partial_{x}A_{1}) && 1-2e\theta(\partial_{x}A_{0}) \\
    1 && - e\theta(\partial_{x}A_{1})&& 0
    && e\theta\left({\cal{F}}_{01}\partial_{x}-\partial_{x}{\cal{F}}_{01}\right)\\
    -e\theta(\partial_{x}A_{1})&& -(1-2e\theta(\partial_{x}A_{0})) &&e\theta
    \left({\cal{F}}_{01}\partial_{x}+2\partial_{x}{\cal{F}}_{01}\right) && 0 \\
  \end{array}
\right)\delta(x-x'),\nonumber\\
\end{eqnarray}
that leads to the Poisson brackets of the bosonic fields
\begin{eqnarray}\label{P17}
\{z^{a}(t;x),z^{b}(t;x')\}=W^{ab}(t;x,x')+{\cal{O}}(e^{2},\theta^{3}),
\end{eqnarray}
with $z^{a},z^{b}\in(A^{0}, A^{1},\dot{A}^{0},\dot{A}^{1})$ and
$W^{ab}(t;x,x')$ given in (\ref{P16}).
\subsection{Modified Poisson Brackets of Fermionic Fields}
\par\noindent
To determine the Poisson brackets corresponding to the fermionic
fields $\psi$ and $\bar{\psi}$, let us consider first the fermionic
Lagrangian density up to ${\cal{O}}(e^{2},\theta^{3})$, which is
given by
\begin{eqnarray}\label{P18}
{\cal
L}_{f}&=&i\bar{\psi}\gamma^{\mu}\partial_{\mu}\psi-e\bar\psi\gamma^{\lambda}\psi
A_{\lambda}
+\frac{ie\theta}{2}\dot{\bar\psi}\gamma^{\lambda}\partial_{x}\psi
A_{\lambda}-\frac{ie\theta}{2}\partial_{x}\bar\psi\gamma^{\lambda}\dot{\psi}
A_{\lambda}+
\frac{e\theta^{2}}{8}\ddot{\bar\psi}\gamma^{\lambda}\partial^{2}_{x}\psi
A_{\lambda}\nonumber\\
&&+
\frac{e\theta^{2}}{8}\partial^{2}_{x}\bar\psi\gamma^{\lambda}\ddot{\psi}
A_{\lambda}-
\frac{e\theta^{2}}{4}\partial_{x}\dot{\bar\psi}\gamma^{\lambda}
\partial_{x}\dot{\psi}A_{\lambda}+{\cal{O}}(e^{2},\theta^{3}).
\end{eqnarray}
In (\ref{P18}) higher order time derivatives, denoted by dots,
appear only in the interaction part of the Lagrangian density. In
what follows, we will determine the modified fermionic Poisson
brackets using the general definition of the fermionic part of the
symplectic two form $\Omega(t)$ from (\ref{U7}), i.e.
\begin{eqnarray*}
\Omega_{f}(t)=\sum\limits_{m=0}^{1}\int
dx\left(X^{(m)}(t;x)-Y^{(m)}(t;x)\right),
\end{eqnarray*}
with $X^{(m)}$ and $Y^{(m)}$ given in (\ref{U8}). Equivalently, in
$1+1$ dimensions,
\begin{eqnarray*}
\Omega_{f}(t)=\frac{1}{2}\int dx\ dx'\
\left(W_{\bar{\psi}\psi}(t;x,x')~d\bar{\psi}(t;x)\wedge~d\psi(t;x)+W_{\psi\bar{\psi}}(t;x,x')d\psi(t;x)\wedge~
d\bar{\psi}(t;x')\right),
\end{eqnarray*}
from (\ref{U13}) can be used. The modification will be performed by
replacing the higher order time derivatives by the corresponding
space derivatives. To do this, we will first use the equation of
motion from (\ref{P5}) up to order ${\cal{O}}(e)$
\begin{eqnarray}\label{P19}
\gamma^{\mu}\partial_{\mu}\psi\approx 0,\qquad &\mbox{leading
to}&\qquad
\dot{\psi}=-\gamma^{5}\partial_{x}\psi,\nonumber\\
\partial_{\mu}\bar{\psi}\gamma^{\mu}\approx 0,\qquad
&\mbox{leading to}&\qquad \dot{\bar{\psi}}=
\partial_{x}\bar{\psi}\gamma^{5},
\end{eqnarray}
where in two dimensions $\gamma^{5}=\gamma^{0}\gamma^{1}$. Then,
using (\ref{P19}), higher order time derivatives acting on $\psi$
and $\bar{\psi}$ can be replaced by the corresponding higher order
space derivatives as,\footnote{See also (\ref{U20}) for a general
$D+1$ dimensional case.}
\begin{eqnarray}
\partial^{n}_{0}\psi_{\alpha}&=&\left\{
\begin{array}{rclcrcl}
n&=&2p,&\qquad&\partial_{0}^{2p}\psi_{\alpha}&=&\partial_{x}^{2p}\psi_{\alpha},\\
n&=&2p+1,&\qquad&\partial_{0}^{2p+1}\psi_{\alpha}&=&\left(-\gamma^{5}\right)_{\alpha\beta}
\partial_{x}^{2p+1}\psi^{\beta},
\end{array}
\right. \label{P20}\\
\partial^{n}_{0}\bar{\psi}_{\alpha}&=&\left\{
\begin{array}{rclcrcl}
n&=&2p,&\qquad&\partial_{0}^{2p}\bar{\psi}_{\alpha}&=&\partial_{x}^{2p}\bar{\psi}_{\alpha},\\
n&=&2p+1,&\qquad&\partial_{0}^{2p+1}\bar{\psi}_{\alpha}&=&\partial_{x}^{2p+1}
\bar{\psi}^{\beta}\left(\gamma^{5}\right)_{\beta\alpha}.
\end{array}
\right. \label{P21}
\end{eqnarray}
Using the general expressions of the fermionic momenta
$\Pi_{\psi^{(m)}}$ and $\Pi_{\bar{\psi}^{(m)}}$ from (\ref{P10}) and
replacing higher order time derivatives with the corresponding
higher order space derivatives using (\ref{P20}) and (\ref{P21}), we
arrive at the modified momenta,
\begin{eqnarray}\label{P22}
\hat{\Pi}_{\psi}^{\beta}&=&
i(\gamma^{0})^{\alpha\beta}\bar\psi_{\alpha}-\frac{ie\theta}{2}\partial_{x}\bar\psi_{\alpha}
(\gamma^{\lambda})^{\alpha\beta}A_{\lambda}-
\frac{e\theta^{2}}{8}\partial^{2}_{x}\bar\psi_{\alpha}\left((\gamma^{\lambda}\gamma^{5})
^{\alpha\beta}\partial_{x}A_{\lambda}
+(\gamma^{\lambda})^{\alpha\beta}\partial_{0}A_{\lambda}\right)+{\cal{O}}(e^{2}, \theta^{3}),\nonumber\\
\hat{\Pi}_{\bar{\psi}}^{\alpha}&=&
\frac{ie\theta}{2}(\gamma^{\lambda})^{\alpha\beta}\partial_{x}\psi_{\beta}A_{\lambda}-
\frac{e\theta^{2}}{8}\left((\gamma^{\lambda}\gamma^{5})^{\alpha\beta}\partial_{x}A_{\lambda}
+(\gamma^{\lambda})^{\alpha\beta}\partial_{0}A_{\lambda}\right)
\partial^{2}_{x}\psi_{\beta}+{\cal{O}}(e^{2}, \theta^{3}),\nonumber\\
\hat{\Pi}_{\dot\psi}^{\beta}&=&\frac{e\theta^{2}}{8}\partial^{2}_{x}\bar\psi_{\alpha}
(\gamma^{\lambda})^{\alpha\beta}A_{\lambda}+{\cal{O}}(e^{2}, \theta^{3}),\nonumber\\
\hat{\Pi}_{\dot{\bar{\psi}}}^{\alpha}&=&\frac{e\theta^{2}}{8}A_{\lambda}
(\gamma^{\lambda})^{\alpha\beta}\partial^{2}_{x}\psi_{\beta}+{\cal{O}}(e^{2},
\theta^{3}).
\end{eqnarray}
Using at this stage (\ref{P22}), the modified $X^{(m)},Y^{(m)}$ with
$m=0,1$ from (\ref{U8}), are explicitly given by
\begin{eqnarray}\label{P23}
\hat{X}^{(0)}&=&d\bar\psi
\left(i\gamma^{0}-\frac{ie\theta}{2}\overleftarrow{\partial}_{x}(\gamma^{\lambda}A_{\lambda})-
\frac{e\theta^{2}}{8}\overleftarrow{\partial}^{2}_{x}\big[\gamma^{\lambda}\gamma^{5}\partial_{x}A_{\lambda}
+\gamma^{\lambda}\partial_{0}A_{\lambda}\big]\right) \wedge
d\psi+{\cal{O}}(e^{2},\theta^{3}),\nonumber\\
\hat{Y}^{(0)}&=&d\bar\psi\wedge
\left(\frac{ie\theta}{2}(\gamma^{\lambda}A_{\lambda})\overrightarrow{\partial}_{x}-
\frac{e\theta^{2}}{8}\big[\gamma^{\lambda}\gamma^{5}\partial_{x}A_{\lambda}
+\gamma^{\lambda}\partial_{0}A_{\lambda}\big]\overrightarrow{\partial}^{2}_{x}\right)
d\psi+{\cal{O}}(e^{2},\theta^{3}),\nonumber\\
\hat{X}^{(1)}&=&d\bar\psi
\left(\frac{e\theta^{2}}{8}\overleftarrow{\partial}^{2}_{x}(\gamma^{\lambda}A_{\lambda})\right)
\wedge d\dot\psi= d\bar\psi
\left(-\frac{e\theta^{2}}{8}\overleftarrow{\partial}^{2}_{x}(\gamma^{\lambda}\gamma^{5}A_{\lambda})\right)
\wedge
d(\partial_{x}\psi)+{\cal{O}}(e^{2},\theta^{3}),\nonumber\\
\hat{Y}^{(1)}&=&d\dot{\bar\psi}\wedge
\left(\frac{e\theta^{2}}{8}(\gamma^{\lambda}A_{\lambda})\overrightarrow{\partial}^{2}_{x}\right)
d\psi=d(\partial_{x}{\bar\psi})\wedge
\left(-\frac{e\theta^{2}}{8}(\gamma^{\lambda}\gamma^{5}
A_{\lambda})\overrightarrow{\partial}^{2}_{x}\right)d\psi+{\cal{O}}(e^{2},\theta^{3}).
\end{eqnarray}
Combining these results, the modified coefficients of the fermionic
part of symplectic two form $\hat{W}_{\psi\bar{\psi}}$ and
$\hat{W}_{\bar{\psi}\psi}$ from (\ref{U15}) can be determined. As we
have also mentioned in Sect. II, in $\hat{W}_{\psi\bar{\psi}}$ all
derivatives act on $\psi$, whereas in $\hat{W}_{\bar{\psi}\psi}$
they act on $\bar\psi$. After performing appropriate partial
differentiations in (\ref{P23}) and neglecting the resulting surface
terms, we arrive at
\begin{eqnarray}\label{P24}
\hat{W}_{\bar{\psi}\psi}(t;x,x')&=&\left\{i\gamma^{0}+e\left(\frac{i\theta}{2}(\gamma^{\lambda}\partial_{x}
A_{\lambda})+\frac{\theta^{2}}{8}
(\gamma^{\lambda}\gamma^{5}\partial_{x}^{3}A_{\lambda})+\frac{3\theta^{2}}{8}
(\gamma^{\lambda}\gamma^{5}\partial_{x}^{2}A_{\lambda})
\partial_{x}+
\frac{\theta^{2}}{8}(\gamma^{\lambda}\partial_{x}^{2}\partial_{0}A_{\lambda})\right.\right.\nonumber\\
&&\left.\left. +\frac{\theta^{2}}{4}(\gamma^{\lambda}
\partial_{x}\partial_{0}A_{\lambda})\partial_{x} +
\frac{3\theta^{2}}{8}(\gamma^{\lambda}\gamma^{5}
\partial_{x}A_{\lambda})\partial_{x}^{2}
+\frac{\theta^{2}}{4}(\gamma^{\lambda}\gamma^{5}A_{\lambda})\partial^{3}_{x}\right)\right\}\delta(x-x')
+{\cal{O}}(e^{2},\theta^{3}), \nonumber\\
\hat{W}_{\psi\bar{\psi}}(t;x,x')&=&\left\{i\gamma^{0}+e\left(\frac{i\theta}{2}(\gamma^{\lambda}\partial_{x}
A_{\lambda})
-\frac{\theta^{2}}{8}(\gamma^{\lambda}\partial_{x}^{2}\partial_{0}A_{\lambda})
-\frac{\theta^{2}}{8}(\gamma^{\lambda}\gamma^{5}\partial_{x}^{3}A_{\lambda})
-\frac{\theta^{2}}{4}(\gamma^{\lambda}\partial_{x}\partial_{0}A_{\lambda})\partial_{x}\right.\right.
\nonumber\\
&&\left.\left.
-\frac{3\theta^{2}}{8}(\gamma^{\lambda}\gamma^{5}\partial_{x}^{2}A_{\lambda})\partial_{x}
-\frac{3\theta^{2}}{8}(\gamma^{\lambda}\gamma^{5}
\partial_{x}A_{\lambda})\partial_{x}^{2}-\frac{\theta^{2}}{4}(\gamma^{\lambda}
\gamma^{5}A_{\lambda})\partial_{x}^{3}\right)\right\}\delta(x-x')+{\cal{O}}(e^{2},\theta^{3}).
\nonumber\\
\end{eqnarray}
They are elements of the matrix
\begin{eqnarray*}
\hat{W}_{ij}=\left(
\begin{array}{cc}
0&\hat{W}_{\psi\bar{\psi}}\\
\hat{W}_{\bar{\psi}\psi}&0
\end{array}
\right),
\end{eqnarray*}
with $(i,j)\in (\psi,\bar{\psi})$, whose inverse leads, similar to
the bosonic case, to the modified fermionic Poisson brackets up to
${\cal{O}}(e^{2},\theta^{3})$
\begin{eqnarray}\label{P25}
\{\psi(t;x),\bar{\psi}(t;x')\}_{\mbox{\tiny{PB}}}&=&\hat{W}^{{\psi}\bar{\psi}}(t;x,x')
\nonumber\\
&\simeq&\left\{-i\gamma^{0}+e\left(
\frac{i\theta}{2}(\Gamma^{\lambda}\partial_{x}A_{\lambda})
+\frac{\theta^{2}}{8}(\Gamma^{\lambda}\partial_{x}^{2}\partial_{0}A_{\lambda})
-\frac{\theta^{2}}{8}(\Gamma^{\lambda}\gamma^{5}\partial_{x}^{3}A_{\lambda})
+\frac{\theta^{2}}{4}(\Gamma^{\lambda}\partial_{x}\partial_{0}A_{\lambda})\partial_{x}
\right.\right.
\nonumber\\
&&\left.\left.
-\frac{3\theta^{2}}{8}(\Gamma^{\lambda}\gamma^{5}\partial_{x}^{2}A_{\lambda})\partial_{x}
-\frac{3\theta^{2}}{8}(\Gamma^{\lambda} \gamma^{5}
\partial_{x}A_{\lambda})\partial_{x}^{2}
-\frac{\theta^{2}}{4}(\Gamma^{\lambda}\gamma^{5}A_{\lambda})\partial^{3}_{x}\right)\right\}\delta(x-x')\nonumber\\
\{\bar{\psi}(t;x),\psi(t;x')\}_{\mbox{\tiny{PB}}}&=&\hat{W}^{\bar{\psi}{\psi}}(t;x,x')\nonumber\\
&\simeq&\left\{-i\gamma^{0}+e\left(\frac{i\theta}{2}(\Gamma^{\lambda}\partial_{x}A_{\lambda})
-\frac{\theta^{2}}{8}(\Gamma^{\lambda}\partial_{x}^{2}\partial_{0}A_{\lambda})
+\frac{\theta^{2}}{8}(\Gamma^{\lambda}\gamma^{5}\partial_{x}^{3}A_{\lambda})
-\frac{\theta^{2}}{4}(\Gamma^{\lambda}\partial_{x}\partial_{0}A_{\lambda})\partial_{x}
\right.\right.\nonumber\\
&&\left.\left.
+\frac{3\theta^{2}}{8}(\Gamma^{\lambda}\gamma^{5}\partial_{x}^{2}A_{\lambda})\partial_{x}
+\frac{3\theta^{2}}{8}(\Gamma^{\lambda}\gamma^{5}
\partial_{x}A_{\lambda})\partial_{x}^{2}
+\frac{\theta^{2}}{4}
(\Gamma^{\lambda}\gamma^{5}A_{\lambda})\partial_{x}^{3}\right)\right\}\delta(x-x'),
\end{eqnarray}
and
$\{\psi(t;x),\psi(t;x')\}_{\mbox{\tiny{PB}}}=\{\bar{\psi}(t;x),\bar\psi(t;x')\}_{\mbox{\tiny{PB}}}=0$.
In (\ref{P25}), $\Gamma^{\lambda}$ is defined by
$\Gamma^{\lambda}\equiv
\gamma_{0}\gamma^{\lambda}\gamma_{0}=2g^{\lambda
0}\gamma^{0}-\gamma^{\lambda}$. Note that the modified matrix
elements $\hat{W}_{\psi\bar{\psi}}$ and $\hat{W}_{\bar{\psi}\psi}$
from (\ref{P24}) as well as their inverse operators
$\hat{W}^{\bar{\psi}\psi}$ and $\hat{W}^{\psi\bar{\psi}}$ from
(\ref{P25}) satisfy the relation (\ref{U14}) and (\ref{U18}),
respectively.
\section{Effective Lagrangian density of two-dimensional space-time
noncommutative QED}
\setcounter{equation}{0}
\par\noindent
In the perturbative approach introduced in \cite{ho-1, ho-2,
eliezer}, one redefines at this stage the field variables
$\psi,\bar{\psi}$ and $A_{\mu}$ so that the Poisson brackets of the
redefined fields and their corresponding conjugate momenta are order
by order the same as in the ordinary commutative theory consisting
of first order time derivatives. In Sect. II, denoting the modified
fields by $\tilde{\psi}, \tilde{\bar{\psi}}$ and $\tilde{A}_{\mu}$,
we have performed a perturbative expansion for a generic theory
described by (\ref{U1})-(\ref{U2}) up to second order in the
coupling constant $e$ and arrived at the following modified Poisson
brackets
\begin{eqnarray*}
\{\tilde{A}_{\mu}(t;x),
\tilde{\Pi}_{\nu}(t;x')\}_{\mbox{\tiny{PB}}}\approx
g_{\mu\nu}\delta(x-x'), \qquad \mbox{and}\qquad
\{\tilde{\bar{\psi}}(t;x),\tilde{\psi}(t;x')\}_{\mbox{\tiny{PB}}}\approx
-i\gamma^{0}\delta(x-x'),
\end{eqnarray*}
[see (\ref{U33}) and (\ref{U37})]. In this section, we will follow
the same method and will first redefine the fermionic and bosonic
field variables for two-dimensional space-time noncommutative QED
described by (\ref{P11}), the bosonic part, and (\ref{P18}), the
fermionic part. Eventually, the redefined fields and the
corresponding conjugate momenta will be used to derive an
appropriate effective Lagrangian density up to
${\cal{O}}(e^{2},\theta^{3})$. To this purpose, we will consider the
fermionic and bosonic Poisson parts separately.
\subsection{The Bosonic Part}
\par\noindent
The Lagrangian density of two-dimensional space-time noncommutative
QED up to ${\cal{O}}(e^{2},\theta^{3})$ is given in (\ref{P11}). To
determine the effective Lagrangian, we will first modify the gauge
field $A_{\mu}$ and its corresponding conjugate momentum $\Pi_{\mu}$
using the general Ansatz (\ref{U30})
\begin{eqnarray*}
\tilde{A}_{\mu}\equiv A_{\mu}-e\xi^{\dot A}_{\mu},&\qquad&
\tilde{\Pi}_{\mu}\equiv - \dot A_{\mu}-e\xi_{\mu}^{A}.
\end{eqnarray*}
Since the Lagrangian (\ref{P11}) include only  first order time
derivative in the order ${\cal{O}}(e^{2},\theta^{3})$, we set
$\xi_{\mu}^{\dot{A}}=0$ for $\mu=0,1$. Using further $\Pi_{\mu}^{A}$
from (\ref{P13}), we get
\begin{eqnarray}\label{B1}
\xi_{0}^{A}&=&\theta\left(\partial_{x}A_{1}\right)(\dot{A}_{1}-\partial_{x}A_{0})+{\cal{O}}(e^{2},\theta^{3}),\nonumber\\
\xi_{1}^{A}&=&\theta\left(2 \dot
A_{1}(\partial_{x}A_{0})-(\partial_{x}A_{0})^{2}-\dot
A_{0}(\partial_{x}A_{1})\right)+{\cal{O}}(e^{2},\theta^{3}).
\end{eqnarray}
It can be easily shown that, the modified bosonic field and its
canonical conjugate momentum satisfy the canonical Poisson bracket
up to order ${\cal{O}}(e^{2},\theta^{3})$
\begin{eqnarray*}
\{\tilde{A}_{\mu}(t;x),
\tilde{\Pi}_{\nu}(t;x')\}_{\mbox{\tiny{PB}}}\approx
g_{\mu\nu}\delta(x-x').
\end{eqnarray*}
In the next step, we will determine the effective bosonic Lagrangian
density in terms of the modified fields. Performing a Legendre
transformation of the Lagrangian density ${\cal{L}}_{g}$ from
(\ref{P11}), the original Hamiltonian is given by
\begin{eqnarray}\label{B2}
{\cal{H}}_{g}(A_{0},A_{1},\dot{A}_{0},\dot{A}_{1})&=&\Pi^{\mu}_{A}\dot
A _{\mu}+\Pi^{\mu}_{\dot A}\ddot{A}_{\mu}-{\cal L}_{g}
\nonumber\\
&=&-\frac{1}{2}\dot A_{0}^{2}+\frac{1}{2}\dot
A_{1}^{2}+\frac{1}{2}(\partial_{x}A_{1})^{2}-\frac{1}{2}(\partial_{x}A_{0})^{2}+e\theta\left((\partial_{x}A_{0})\dot
A_{1}^{2}-(\partial_{x}A_{1})\dot A_{0}\dot
A_{1}\right)\nonumber\\
&&+{\cal O}(e^{2},\theta^{3}).
\end{eqnarray}
To determine the Hamiltonian ${\cal{H}}_{g}$ in terms of the
modified variables $\tilde{A}_{\mu}$ and $\tilde{\Pi}_{\mu}$, we
will use
\begin{eqnarray}\label{B3}
\tilde{\Pi}_{0}&=&-\dot A_{0}-e\theta(\partial_{x}A_{1})(\dot
A_{1}-\partial_{x}A_{0}),\nonumber\\
\tilde{\Pi}_{1}&=&-\dot A_{1}-e\theta\left(2\dot
A_{1}(\partial_{x}A_{0})-(\partial_{x}A_{0})^{2}-\dot
A_{0}(\partial_{x}A_{1})\right),
\end{eqnarray}
and determine $\dot{A}_{\mu}, \mu=0,1$ in terms of
$\partial_{x}A_{1}=\partial_{x}\tilde{A}_{1}$ and
$\tilde{\Pi}_{\mu}$. We arrive at
\begin{eqnarray}\label{B4}
\dot A _{0}&=&-\tilde{\Pi}_{0}+e\theta\bigg[
(\partial_{x}\tilde{A}_{1})\tilde{\Pi}_{1}+(\partial_{x}\tilde{A}_{0})(\partial_{x}{\tilde{A}}_{1})\bigg],\nonumber\\
\dot
A_{1}&=&-\tilde{\Pi}_{1}+e\theta\left((\partial_{x}\tilde{A}_{0})^{2}+2(\partial_{x}\tilde{A}_{0})\tilde{\Pi}_{1}
-(\partial_{x}\tilde{A}_{1})\tilde{\Pi}_{0}\right).
\end{eqnarray}
Plugging (\ref{B4}) in the Hamiltonian ${\cal{H}}_{g}$ from
(\ref{B2}), we arrive at the modified Hamiltonian
\begin{eqnarray}\label{B5}
\tilde{{\cal
H}}_{g}(\tilde{A}_{\mu},\tilde{\Pi}_{\mu})&=&\frac{1}{2}\tilde{\Pi}_{1}^{2}-\frac{1}{2}\tilde{\Pi}_{0}^{2}+\frac{1}{2}(\partial_{x}\tilde{A}_{1})^{2}-\frac{1}{2}(\partial_{x}\tilde{A}_{0})^{2}
\nonumber\\
&&+e\theta\bigg[(\partial_{x}\tilde{A}_{0})(\partial_{x}\tilde{A}_{1})\tilde{\Pi}_{0}-(\partial_{x}\tilde{A}_{0})\tilde{\Pi}_{1}^{2}-(\partial_{x}\tilde{A}_{0})^{2}\tilde{\Pi}_{1}+(\partial_{x}\tilde{A}_{1})
\tilde{\Pi}_{0}\tilde{\Pi}_{1}\bigg]+{\cal
O}((e\theta)^{2}).\nonumber\\
\end{eqnarray}
The effective Lagrangian of the modified gauge fields is then
defined as
\begin{eqnarray}\label{B6}
\tilde{{\cal L}}_{g}\equiv\tilde{\Pi}^{\mu}
{{\dot{\tilde{A}}}}_{\mu}-\tilde{{\cal
H}}_{g}(\tilde{A}_{\mu},\tilde{\Pi}_{\mu}).
\end{eqnarray}
To determine $\dot{\tilde{A}}_{\mu}$ we use the Heisenberg equation
of motion
\begin{eqnarray}\label{B7}
\dot{\tilde{A}}_{\mu}=\{\tilde{A}_{\mu},\tilde{{\cal H}}_{g}\}.
\end{eqnarray}
Using further the modified equal-time Poisson brackets
$\{\tilde{A}_{\mu}(t;x),
\tilde{\Pi}_{\nu}(t;x')\}_{\mbox{\tiny{PB}}}=
g_{\mu\nu}\delta(x-x')+{\cal{O}}(e^{2},\theta^{3})$ from
(\ref{U33}), it can easily be shown that
$\dot{\tilde{A}}_{\mu}\approx \dot{A}_{\mu}$ up to order
${\cal{O}}(e^{2},\theta^{3})$. Plugging this result back in
(\ref{B6}), the effective Lagrangian density of the gauge fields is
given by
\begin{eqnarray}\label{B8}
\tilde{{\cal
L}}_{g}&=&\frac{1}{2}\dot{\tilde{A}}_{1}^{2}-\frac{1}{2}\dot{\tilde{A}}_{0}^{2}-
\frac{1}{2}(\partial_{x}\tilde{A}_{1})^{2}+\frac{1}{2}(\partial_{x}\tilde{A}_{0})^{2}
+e\theta
\bigg[(\partial_{x}\tilde{A}_{0})\dot{\tilde{A}}_{1}-(\partial_{x}\tilde{A}_{1})\dot{\tilde{A}}_{0}
\bigg]\tilde{{\cal
F}}_{01}+{\cal{O}}((e\theta)^{2})\nonumber\\
&=&-\frac{1}{2}(\partial_{\mu}\tilde{A}_{\nu})(\partial^{\mu}\tilde{A}^{\nu})+\frac{e\theta}{2}\epsilon^{\alpha\beta}\partial_{\alpha}\tilde{A}^{\mu}\partial_{\beta}\tilde{A}^{\nu}\tilde{{\cal
F}}_{\mu\nu}+{\cal{O}}((e\theta)^{2}),
\end{eqnarray}
where $\tilde{\cal{F}}_{\mu\nu}\equiv
\partial_{\mu}\tilde{A}_{\nu}-\partial_{\nu}\tilde{A}_{\mu}$. Note that the $\theta$ dependent part of (\ref{B8}) is, as expected, exactly the same as
the original Lagrangian ${\cal{L}}_{g}$ from (\ref{P11}).
\subsection{The Fermionic Part}
\par\noindent
First let us modify the fermionic fields using the Ansatz
\begin{eqnarray}\label{B9}
\qquad\tilde{\psi}=(1+ie\xi_{\bar{\psi}}\gamma^{0})\psi,\qquad\mbox{and}\qquad
\tilde{\bar{\psi}}=\bar{\psi}(1-ie\xi_{\psi} \gamma^{0}),
\end{eqnarray}
 where compared to (\ref{U36}),
$\xi_{\bar{\psi}}=-i\overrightarrow{\cal{J}}_{\bar{\psi}}\gamma^{0}$
and $\xi_{\psi}=i\overleftarrow{\cal{J}}_{\psi}\gamma^{0}$. Choosing
$\xi_{\bar\psi}$ and $\xi_{\psi}$ as
\begin{eqnarray}\label{B10}
\xi_{\bar{\psi}}&=&0,\nonumber\\
\xi_{\psi}&=&\frac{i\theta}{2}\gamma^{\lambda}\partial_{x}A_{\lambda}+
\frac{\theta^{2}}{8}\gamma^{\lambda}
\gamma^{5}\partial_{x}^{3}A_{\lambda}+
\frac{\theta^{2}}{8}\gamma^{\lambda}\partial_{x}^{2}\partial_{0}A_{\lambda}+
\frac{\theta^{2}}{4}\gamma^{\lambda}
\overleftarrow{\partial_{x}}(\partial_{x}\partial_{0}A_{\lambda})+
\frac{3\theta^{2}}{8}\gamma^{\lambda} \gamma^{5}
\overleftarrow{\partial_{x}}(\partial_{x}^{2}A_{\lambda})
\nonumber\\
&&+\frac{3\theta^{2}}{8}\gamma^{\lambda}\gamma^{5}
\overleftarrow{\partial_{x}}^{2}(\partial_{x}A_{\lambda})
+\frac{\theta^{2}}{4}\gamma^{\lambda}\gamma^{5}\overleftarrow{\partial_{x}}^{3}A_{\lambda}
+{\cal{O}}(e,\theta^{3}),
\end{eqnarray}
the canonical fermionic Poisson brackets,
\begin{eqnarray*}
\{\tilde{\bar{\psi}}(t;x),\tilde{\psi}(t;x')\}_{\mbox{\tiny{PB}}}\approx
-i\gamma^{0}\delta(x-x'),
\end{eqnarray*}
can be shown to be valid up to ${\cal{O}}(e^{2},\theta^{3})$, as
expected. In what follows, we will determine the fermionic part of
the Hamiltonian density as a function of $\tilde{\psi}$ and
$\tilde{\bar{\psi}}$. The original Hamiltonian in terms of $\psi$
and $\bar{\psi}$ is given by the Legendre transformation of the
Lagrangian density ${\cal{L}}_{f}$ from (\ref{P18})
\begin{eqnarray}\label{B11}
{\cal{H}}_{f}(\psi,\bar{\psi})&=&\Pi_{\psi}^{\beta}\dot{\psi}_{\beta}+\Pi_{\dot{\psi}}^{\beta}\ddot{\psi}_{\beta}+\dot{\bar{\psi}}_{\beta}\Pi_{\bar{\psi}}^{\beta}
+\ddot{\bar{\psi}}_{\beta}\Pi_{\dot{\bar{\psi}}}^{\beta}-{\cal{L}}_{f}(\psi, \bar{\psi})\nonumber\\
&=&
-i\bar{\psi}\gamma^{1}\partial_{x}\psi+e\bar{\psi}\gamma^{\lambda}\psi
A_{\lambda}\nonumber\\
&& +\frac{e\theta^{2}}{8}\partial_{x}\bar{\psi}
\left(\gamma^{\lambda}\partial_{x}A_{\lambda}+\gamma^{\lambda}\gamma^{5}
\partial_{0}A_{\lambda}\right)\partial_{x}^{2}\psi+
\frac{e\theta^{2}}{8}\partial_{x}^{2}\bar{\psi}
\left(\gamma^{\lambda}\partial_{x}A_{\lambda}+\gamma^{\lambda}\gamma^{5}
\partial_{0}A_{\lambda}\right)\partial_{x}\psi\nonumber\\
&&+\frac{e\theta^{2}}{4}\partial_{x}^{2}\bar{\psi}\
\gamma^{\lambda}\partial_{x}^{2}\psi\
A_{\lambda}+{\cal{O}}(e^{2},\theta^{3}),
\end{eqnarray}
To formulate ${\cal{H}}_{f}$ in terms of redefined fields
$\tilde{\psi}$ and $\tilde{\bar{\psi}}$, we will first invert
(\ref{B9}) to get
\begin{eqnarray}\label{B12}
\psi=\tilde{\psi},\qquad\mbox{and}\qquad\bar{\psi}=
\tilde{\bar{\psi}}(1+ie\xi_{\psi}\gamma^{0})+{\cal{O}}(e^{2},\theta^{3}),
\end{eqnarray}
where $\xi_{\bar{\psi}}$ and $\xi_{\psi}$ can be read from
(\ref{B10}). Then replacing (\ref{B12}) in (\ref{B11}), we arrive at
\begin{eqnarray}\label{B13}
\tilde{\cal H}_{f}&\equiv&{\cal
H}_{f}\left(\bar{\psi}(\tilde{\bar{\psi}},A_{\mu}),\psi(\tilde{\psi},A_{\mu})\right)=
-i\tilde{\bar{\psi}}(1+ie\xi_{\psi}\gamma^{0})\gamma^{1}\partial_{x}\tilde{\psi}+
e\tilde{\bar{\psi}}\gamma^{\lambda}\tilde{\psi}
A_{\lambda}\nonumber\\
&&+\frac{e\theta^{2}}{8}\partial_{x}\tilde{\bar{\psi}}\left(\gamma^{\lambda}\partial_{x}
A_{\lambda}+\gamma^{\lambda}\gamma^{5}\partial_{0}A_{\lambda}\right)\partial_{x}^{2}\tilde{\psi}
+\frac{e\theta^{2}}{8}\partial_{x}^{2}\tilde{\bar{\psi}}\left(\gamma^{\lambda}
\partial_{x}A_{\lambda}+\gamma^{\lambda}\gamma^{5}\partial_{0}A_{\lambda}\right)
\partial_{x}\tilde{\psi}\nonumber\\
&&+\frac{e\theta^{2}}{4}
\partial_{x}^{2}\tilde{\bar{\psi}}\gamma^{\lambda}\partial_{x}^{2}\tilde{\psi}
A_{\lambda}+{\cal{O}}(e^{2},\theta^{3}),
\end{eqnarray}
where $\xi_{\psi}$ is given in (\ref{B10}). Due to the special form
of redefined fields in (\ref{B9}) and (\ref{B12}), the above
modified Hamiltonian is not Hermitian. The Hermitian conjugate of
$-i\tilde{\bar{\psi}}\gamma^{1}\partial_{x}\tilde{\psi}$ on the
first line is to be added to $\tilde{\cal{H}}_{f}$ to build a
Hermitian Hamiltonian. We denote it by $ {\cal{C}}\equiv
i\left(\tilde{\bar{\psi}}
\gamma^{1}\partial_{x}\tilde{\psi}\right)^{\dagger}$. Adding
${\cal{C}}$ to $\tilde{\cal{H}}_{f}$ and using the standard
definition
\begin{eqnarray}\label{B14}
\tilde{\cal L}_{f}=\tilde{\cal
L}_{f}(\tilde{\bar{\psi}},\tilde{\psi})=
i\tilde{\bar{\psi}}\gamma^{0}\dot{\tilde{\psi}}-\tilde{\cal H}_{f},
\end{eqnarray}
the modified Lagrangian density $\tilde{\cal{L}}_{f}$ in terms of
the redefined fields is given as
\begin{eqnarray}\label{B15}
\tilde{\cal L}_{f}&=&
i\tilde{\bar{\psi}}\gamma^{\mu}\partial_{\mu}\tilde{\psi}-e\tilde{\bar{\psi}}\gamma^{\lambda}\tilde{\psi}A_{\lambda}\nonumber\\
&&
-i\partial_{x}\bar{\tilde{\psi}}\gamma^{1}\gamma^{0}\tilde{\bar{\psi}}^{\dagger}
-e\left\{\tilde{\bar{\psi}}\left(\frac{i\theta}{2}\gamma^{\lambda}\gamma^{5}\partial_{x}A_{\lambda}\partial_{x}
-\frac{\theta^{2}}{8}\gamma^{\lambda}\gamma^{5}\partial_{x}\partial_{0}A_{\lambda}
\partial_{x}^{2}
+\frac{\theta^{2}}{8}\gamma^{\lambda}\partial_{x}A_{\lambda}
\partial_{x}^{3}\right)\tilde{\psi}\right.\nonumber\\
&&\left.+\partial_{x}\bar{\tilde{\psi}}
\left(-\frac{i\theta}{2}\gamma^{\lambda}\gamma^{5}\partial_{x}A_{\lambda}+
\frac{\theta^{2}}{8}\gamma^{\lambda}
\partial_{x}^{3}A_{\lambda}+\frac{3\theta^{2}}{8}\gamma^{\lambda}
\partial_{x}^{2}A_{\lambda}\partial_{x}
+\frac{\theta^{2}}{8}\gamma^{\lambda}\gamma^{5}\partial_{x}^{2}\partial_{0}A_{\lambda}+
\frac{\theta^{2}}{4}\gamma^{\lambda}\gamma^{5}
\partial_{x}\partial_{0}A_{\lambda}\partial_{x}
\right.\right.\nonumber\\
&&\left.\left.\qquad\qquad
+\frac{\theta^{2}}{4}\gamma^{\lambda}A_{\lambda}\partial^{3}_{x}
+\frac{3\theta^{2}}{8}\gamma^{\lambda}
\partial_{x}A_{\lambda}\partial_{x}^{2}\right)\gamma^{0}\tilde{\bar{\psi}}^{\dagger}\right\}
+{\cal{O}}(e^{2},\theta^{3}).
\end{eqnarray}
To simplify the above modified Lagrangian, we will use the following
relations
\begin{eqnarray}\label{B16}
\bar{\tilde{\psi}}&=&\tilde{\bar{\psi}}(1+ie\xi_{\psi}\gamma^{0})\nonumber\\
&=&\tilde{\bar{\psi}}+ie\tilde{\bar{\psi}}\left(\frac{i\theta}{2}\gamma^{\lambda}\gamma^{0}
\partial_{x}A_{\lambda}+\frac{\theta^{2}}{8}\gamma^{\lambda}\gamma^{0}
\partial_{x}^{2}\partial_{0}A_{\lambda}+
\frac{\theta^{2}}{8}\gamma^{\lambda}
\gamma^{5}\gamma^{0}\partial_{x}^{3}A_{\lambda}+\frac{3\theta^{2}}{8}\gamma^{\lambda}
\gamma^{5}\gamma^{0}
\overleftarrow{\partial_{x}}(\partial_{x}^{2}A_{\lambda})\right.\nonumber\\
&&\qquad\qquad \left.+\frac{\theta^{2}}{4}\gamma^{\lambda}\gamma^{0}
\overleftarrow{\partial_{x}}(\partial_{x}\partial_{0}A_{\lambda})
+\frac{\theta^{2}}{4}\gamma^{\lambda}\gamma^{5}\gamma^{0}
\overleftarrow{\partial^{3}_{x}}A_{\lambda}+\frac{3\theta^{2}}{8}\gamma^{\lambda}\gamma^{5}\gamma^{0}
\overleftarrow{\partial_{x}^{2}}(\partial_{x}A_{\lambda})\right)+{\cal{O}}(e^{2},\theta^{3}),\nonumber\\
\tilde{\bar{\psi}}^{\dagger}&=&
(1+ie\gamma_{0}\xi_{\psi}^{\dagger})\gamma_{0}\tilde{\psi}\nonumber\\
&=&\gamma_{0}\tilde{\psi}-ie\left(\frac{i\theta}{2}\gamma^{\lambda}\partial_{x}A_{\lambda}-
\frac{\theta^{2}}{8}\gamma^{\lambda}
\gamma^{5}\partial_{x}^{3}A_{\lambda}
-\frac{\theta^{2}}{8}\gamma^{\lambda}\partial_{x}^{2}\partial_{0}A_{\lambda}-
\frac{3\theta^{2}}{8}\gamma^{\lambda}
\gamma^{5}\partial_{x}^{2}A_{\lambda}{\partial_{x}}
\right.\nonumber\\
&&\left.\qquad\qquad -\frac{\theta^{2}}{4}\gamma^{\lambda}
\partial_{x}\partial_{0}A_{\lambda}{\partial_{x}}
-\frac{3\theta^{2}}{8}\gamma^{\lambda}\gamma^{5}
\partial_{x}A_{\lambda}{\partial_{x}^{2}}
-\frac{\theta^{2}}{4}\gamma^{\lambda}\gamma^{5}
A_{\lambda}{\partial^{3}_{x}}\right)\tilde{\psi}+{\cal{O}}(e^{2},\theta^{3}),
\end{eqnarray}
where
\begin{eqnarray}\label{B17}
\xi_{\psi}^{\dagger}&=&-\frac{i\theta}{2}\Gamma^{\lambda}\partial_{x}A_{\lambda}+
\frac{\theta^{2}}{8}\gamma^{5}\Gamma^{\lambda}
\partial_{x}^{3}A_{\lambda}+\frac{3\theta^{2}}{8}\gamma^{5}\Gamma^{\lambda}
\partial_{x}^{2}A_{\lambda}\partial_{x}+\frac{\theta^{2}}{8}\Gamma^{\lambda}\gamma^{0}\partial_{x}^{2}\partial_{0}
A_{\lambda} +\frac{\theta^{2}}{4}\Gamma^{\lambda}
\partial_{x}\partial_{0}A_{\lambda}\partial_{x}
\nonumber\\
&&
+\frac{\theta^{2}}{4}\gamma^{5}\Gamma^{\lambda}A_{\lambda}\partial^{3}_{x}
+\frac{3\theta^{2}}{8}\gamma^{5}\Gamma^{\lambda}
\partial_{x}A_{\lambda}\partial_{x}^{2}+{\cal{O}}(e^{2},\theta^{3}),
\end{eqnarray}
with $\Gamma^{\lambda}=\gamma^{0}\gamma^{\lambda}\gamma^{0}$ is
used. We arrive finally at the effective Lagrangian density
including only first order time derivative of bosonic and fermionic
fields
\begin{eqnarray}\label{B18}
\tilde{\cal L}_{f}&=&
\tilde{\bar{\psi}}\left\{i\gamma^{\mu}\partial_{\mu}-e\left(\gamma^{\lambda}A_{\lambda}+\frac{i\theta}{2}
\gamma^{\lambda}\gamma^{5}(\partial_{x}A_{\lambda})\partial_{x}-
\frac{\theta^{2}}{8}\gamma^{\lambda}\gamma^{5}(\partial_{x}\partial_{0}
A_{\lambda})\partial_{x}^{2}+
\frac{\theta^{2}}{8}\gamma^{\lambda}(\partial_{x}A_{\lambda})\partial_{x}^{3}
\right)\right\}\tilde{\psi}\nonumber\\
&&+{\cal{O}}(e^{2},\theta^{3}).
\end{eqnarray}
Combining at this stage the bosonic and fermionic parts of the
Lagrangian density from (\ref{B8}) and (\ref{B18}), we arrive at the
total effective Lagrangian density in term of the redefined fields
$\tilde{\psi},\tilde{\bar{\psi}}$ and $\tilde{A}_{\mu}$
\begin{eqnarray}\label{B19}
\lefteqn{
\widetilde{\cal{L}}[\tilde{\psi},\tilde{\bar{\psi}},\tilde{A}_{\mu}]=}\nonumber\\
&=&
-\frac{1}{2}(\partial_{\mu}\tilde{A}_{\nu})(\partial^{\mu}\tilde{A}^{\nu})+\frac{e\theta}{2}\epsilon^{\alpha\beta}\partial_{\alpha}\tilde{A}^{\mu}\partial_{\beta}\tilde{A}^{\nu}\tilde{{\cal
F}}_{\mu\nu}\nonumber\\
&&+\tilde{\bar{\psi}}\left\{i\gamma^{\mu}\partial_{\mu}-e\left(\gamma^{\lambda}A_{\lambda}+\frac{i\theta}{2}
\gamma^{\lambda}\gamma^{5}(\partial_{x}\tilde{A}_{\lambda})\partial_{x}-
\frac{\theta^{2}}{8}\gamma^{\lambda}\gamma^{5}(\partial_{x}\partial_{0}
\tilde{A}_{\lambda})\partial_{x}^{2}+
\frac{\theta^{2}}{8}\gamma^{\lambda}(\partial_{x}\tilde{A}_{\lambda})\partial_{x}^{3}
\right)\right\}\tilde{\psi}\nonumber\\
&&+{\cal{O}}(e^{2},\theta^{3}).
\end{eqnarray} The final Lagrangian
includes first order time derivative and higher order space
derivatives. The Lorentz covariance of the modified Lagrangian is
broken by the procedure of perturbative quantization, where higher
order time derivatives are replaced by corresponding space
derivatives. The above Lagrangian can be nevertheless regarded as
the starting point for further perturbative and nonperturbative
study of two-dimensional space-time noncommutative QED up to order
${\cal{O}}(e^{2},\theta^{3})$.
\section{The Algebra of currents of the modified noncommutative theory}
\setcounter{equation}{0}\par\noindent In this section, as a possible
application of our previous results, we will use the Dirac brackets
of the modified fermionic and bosonic fields to determine the
current algebra of global $U(1)$ vector current (\ref{P6})
\begin{eqnarray*}
J^{\mu}(x)=-\psi_{\beta}(x)\star\bar\psi_{\alpha}(x)
(\gamma^{\mu})^{\alpha\beta},
\end{eqnarray*}
corresponding to the original two-dimensional noncommutative QED
described by (\ref{P4}). Using the definition of the $\star$-product
from (\ref{P1}), $J^{\mu}(x)$ can be written as
\begin{eqnarray}\label{S1}
J^{\mu}(x)=\bar\psi \gamma^{\mu}\psi+\frac{i\theta}{2}
\epsilon^{\rho\sigma}\partial_{\sigma}\bar\psi\gamma^{\mu}\partial_{\rho}\psi-
\frac{\theta^{2}}{8}\epsilon^{\rho\sigma}\epsilon^{\lambda\eta}
\partial_{\sigma}\partial_{\eta}\bar\psi\gamma^{\mu}\partial_{\rho}\partial_{\lambda}\psi+{\cal{O}}(\theta^{3}).
\end{eqnarray}
Replacing $\psi$ and $\bar{\psi}$ with the modified fermionic fields
$\tilde{\psi}$ and $\tilde{\bar{\psi}}$ using the relations
(\ref{B12}), we arrive after some lengthy but straightforward
manipulations at the corresponding modified currents
\begin{eqnarray}
\tilde{J}^{0}(x)&=&\tilde{\bar\psi}\gamma^{0}\tilde{\psi}
+ie\tilde{\bar\psi}\left(\frac{i\theta}{2}\gamma^{\lambda}\partial_{x}A_{\lambda}+
\frac{\theta^{2}}{8}\gamma^{\lambda}
\gamma^{5}\partial_{x}^{3}A_{\lambda}+\frac{\theta^{2}}{8}\gamma^{\lambda}
\partial_{x}^{2}\partial_{0}A_{\lambda}
+\frac{3\theta^{2}}{8}\gamma^{\lambda} \gamma^{5}
\overleftarrow{\partial_{x}}(\partial_{x}^{2}A_{\lambda})
\right.\nonumber\\
&&\left. +\frac{\theta^{2}}{4}\gamma^{\lambda}
\overleftarrow{\partial_{x}}(\partial_{x}\partial_{0}A_{\lambda})
+\frac{3\theta^{2}}{8}\gamma^{\lambda}\gamma^{5}
\overleftarrow{\partial_{x}^{2}}(\partial_{x}A_{\lambda})+
\frac{\theta^{2}}{4}\gamma^{\lambda}\gamma^{5}\overleftarrow{\partial^{3}_{x}}A_{\lambda}
\right)\tilde{\psi}
+\frac{i\theta}{2}\epsilon^{\rho\sigma}\partial_{\sigma}\tilde{\bar\psi}\gamma^{0}
\partial_{\rho}\tilde{\psi}
\nonumber\\
&&-\frac{\theta^{2}}{8}\epsilon^{\rho\sigma}\epsilon^{\lambda\eta}
\partial_{\sigma}\partial_{\eta}\tilde{\bar\psi}\gamma^{0}
\partial_{\rho}\partial_{\lambda}\tilde{\psi}-\frac{ie\theta^{2}}{4}\epsilon^{\rho\sigma}
\partial_{\sigma}(
\tilde{\bar\psi}\partial_{x}A_{\lambda})\gamma^{\lambda}\partial_{\rho}\tilde{\psi}
+{\cal O}(e^{2},\theta^{3}), \label{S2}\\
\tilde{J}^{1}(x)&=&\tilde{\bar\psi}\gamma^{1}\tilde{\psi}+ie\tilde{\bar\psi}
\left(\frac{i\theta}{2}\gamma^{\lambda}\gamma^{5}\partial_{x}A_{\lambda}+
\frac{\theta^{2}}{8}\gamma^{\lambda}
\partial_{x}^{3}A_{\lambda}+\frac{\theta^{2}}{8}
\gamma^{\lambda}\gamma^{5}\partial_{x}^{2}\partial_{0}A_{\lambda}
+\frac{3\theta^{2}}{8}\gamma^{\lambda}
\overleftarrow{\partial_{x}}(\partial_{x}^{2}A_{\lambda})
\right.\nonumber\\
&&\left. +\frac{\theta^{2}}{4}\gamma^{\lambda}\gamma^{5}
\overleftarrow{\partial_{x}}(\partial_{x}\partial_{0}A_{\lambda})
+\frac{3\theta^{2}}{8}\gamma^{\lambda}
\overleftarrow{\partial_{x}^{2}}(\partial_{x}A_{\lambda})+\frac{\theta^{2}}{4}
\gamma^{\lambda}\overleftarrow{\partial^{3}_{x}}A_{\lambda}\right)\tilde{\psi}
+
\frac{i\theta}{2}\epsilon^{\rho\sigma}\partial_{\sigma}\tilde{\bar\psi}\gamma^{1}
\partial_{\rho}\tilde{\psi}
\nonumber\\
&& -\frac{\theta^{2}}{8}\epsilon^{\rho\sigma}\epsilon^{\lambda\eta}
\partial_{\sigma}\partial_{\eta}\tilde{\bar\psi}\gamma^{1}
\partial_{\rho}\partial_{\lambda}\tilde{\psi}-\frac{ie\theta^{2}}{4}\epsilon^{\rho\sigma}
\partial_{\sigma}(\tilde{\bar\psi}\partial_{x}A_{\lambda})\gamma^{\lambda}\gamma^{5}
\partial_{\rho}\tilde{\psi}+{\cal O}(e^{2},\theta^{3}). \label{S3}
\end{eqnarray}
Using now the equal-time commutation relations (\ref{U38}) between
redefined field operators $\tilde{\psi}$ and $\tilde{\bar{\psi}}$,
we arrive at the following algebra of currents
\begin{eqnarray}
[\tilde{J}^{0}(t;x),\tilde{J}^{1}(t;x')]\bigg|_{\mbox{\tiny{canonical}}}&=&
-i\theta\partial_{0}(\tilde{\bar\psi}\gamma^{1}
\tilde{\psi})\partial_{x}\delta(x-x')
+ie\theta^{2}(\partial_{x}A_{\lambda})\partial_{0}(\tilde{\bar\psi}\gamma^{\lambda}\gamma^{5}\tilde{\psi})
\partial_{x}\delta(x-x')\nonumber\\&&
+\frac{ie\theta^{2}}{4}
(\partial_{0}\partial_{x}A_{\lambda})(\tilde{\bar\psi}\gamma^{\lambda}\gamma^{5}\tilde{\psi})
\partial_{x}\delta(x-x')+{\cal{O}}(e^{2},\theta^{3}),\\ \label{S5}
[\tilde{J}^{0}(t;x),\tilde{J}^{0}(t;x')]\bigg|_{\mbox{\tiny{canonical}}}&=&
[\tilde{J}^{1}(t;x),\tilde{J}^{1}(t;x')]\bigg|_{\mbox{\tiny{canonical}}}
\nonumber\\
&=&-i\theta\partial_{0}(\tilde{\bar\psi}\gamma^{0}\tilde{\psi})\partial_{x}\delta(x-x')
+ie\theta^{2}(\partial_{x}A_{\lambda})\partial_{0}(\tilde{\bar\psi}\gamma^{\lambda}\tilde{\psi})
\partial_{x}\delta(x-x')\nonumber\\&&
+\frac{ie\theta^{2}}{4}
(\partial_{0}\partial_{x}A_{\lambda})(\tilde{\bar\psi}\gamma^{\lambda}\tilde{\psi})\partial_{x}\delta(x-x')
+{\cal{O}}(e^{2},\theta^{3}), \label{S6}
\end{eqnarray}
that contains Schwinger terms on the right hand sides.
\section{Concluding remarks}
\par\noindent
In the first part of the paper, we have presented the general
framework of perturbative quantization for a $D+1$ dimensional
QED-like theory, that includes bosons and fermions whose
interactions are described by terms containing higher order
space-time derivatives. According to general procedure described in
\cite{ho-1, ho-2}, the equations of motion of the original field
theory are used to define time derivatives as a function of space
derivatives in the lowest order of perturbative expansion in the
order of QED coupling constant $e$. Then, the fermionic and bosonic
field variables are appropriately modified, so that they satisfy the
ordinary fundamental Poisson brackets in this first order
approximation. Using the standard Dirac quantization procedure, the
equal-time commutation relations corresponding to fermions and
bosons are determined up to ${\cal{O}}(e^{2})$.
\par
In the second part of the paper, two-dimensional space-time NC-QED
is perturbatively quantized up to ${\cal{O}}(e^{2},\theta^{3})$,
where $\theta$ is the space-time noncommutativity parameter.
Noncommutative field theories, in general, are characterized by a
noncommutative Moyal product that replaces the ordinary product of
functions in commutative field theory. In two dimensions, in
particular, the Moyal product involves an infinite number of
space-time derivatives. Appearing in the interaction part of the
theory, the space-time noncommutativity renders the theory acausal
and inconsistent with conventional Hamiltonian evolution
\cite{susskind}. The S-matrix of the theory is also non-unitary
\cite{mehen}. Different attempts are performed to cure space-time
NC-QED \cite{fredenhagen}, that in two dimensions are by themselves
interesting to study not only because they are the noncommutative
counterpart of the well-known Schwinger model \cite{schwinger}.
\par
Following the procedure of perturbative quantization, we have
determined the Lagrangian density of two-dimensional space-time QED
in terms of modified field variables that satisfy the ordinary
equal-time commutation relations up to
${\cal{O}}(e^{2},\theta^{3})$. In this lowest approximation,
although the Poisson algebra of the bosonic field variables are
modified by terms proportional to the noncommutativity parameter
$\theta$, the bosonic part of the Lagrangian density remains
unchanged. The fermionic part consists of first order time
derivatives and higher order space derivatives of bosonic and
fermionic field variables. The modified Lagrangian density
(\ref{B19}) has lost, due to the special feature of perturbative
quantization, the relativistic covariance of space and time
coordinates. Using the canonical equal-time commutation relations of
the modified field variables, the algebra of global NC-$U_{V}(1)$
currents of the original NC-QED is also determined. In summary, the
modified Lagrangian density (\ref{B19}) can be regarded as the
starting point for further perturbative and nonperturbative study of
two-dimensional space-time noncommutative QED up to order
${\cal{O}}(e^{2},\theta^{3})$, that will be the subject of future
publications.
\section{Acknowledgements}
\par\noindent
The authors thank F. Ardalan for useful discussions. M. Gh. thanks
C. M. Reyes for e-mail correspondence.

\end{document}